\pdfoutput=1
\documentclass[a4paper,11pt]{article}
\usepackage{jcappub}
\usepackage{bm}
\usepackage{booktabs}
\usepackage{xcolor}

\newcommand{\dd}{\mathrm{d}}
\newcommand{\rs}{r_{\!*}}
\newcommand{\Veff}{V_{\ell}^{\pm}}
\newcommand{\bc}{b_{\mathrm{c}}}
\newcommand{\Cwey}{C}
\newcommand{\shpanel}[3]{%
  \begin{minipage}{#1}\centering
    \includegraphics[width=\linewidth]{#2}\\[-1pt]
    {\scriptsize $\bc/M=#3$}%
  \end{minipage}}

\title{Polarization-dependent observational signatures of Weyl-coupled photons around a black hole}

\author[a]{Mao-Yuan Wan,}
\author[a,1]{Chen Wu}
\affiliation[a]{Xingzhi College, Zhejiang Normal University, Jinhua 321004, Zhejiang, China}
\emailAdd{maoyuan.wan.physics@gmail.com}
\emailAdd{wuchenoffd@gmail.com}
\note[1]{Corresponding author.}

\abstract{%
A photon coupled non-minimally to the Weyl tensor propagates differently in its two linear polarization states: the vacuum around a black hole becomes a birefringent medium.
On a Schwarzschild background the coupling $\alpha$ splits the photon sphere and the critical impact parameter of the two polarizations,
producing a double shadow and a polarization-dependent photon ring;
the two shadow edges separate by $62\%$ of the shadow radius at $\alpha/M^2=0.75$.
We compute a unified set of wave-optical and geometric-optical observables for both polarizations across a range of $\alpha$:
greybody factors, absorption cross sections, differential cross sections and their glories, a backward birefringence signal, shadows, photon rings, and null trajectories.
The two-polarization scattering matrices are extracted with a Riccati--Hankel matching scheme that keeps the phase-sensitive observables reliable,
and the geometric-optics chain is cross-validated with three independent ray tracers.
The cleanest diagnostic is the backward signal: it is protected by parity and vanishes identically at $\alpha=0$.
Laboratory bounds on $\alpha$ render these effects unobservable for supermassive black holes;
order-unity couplings would require a black hole of primordial mass.}


\begin{document}
\maketitle

\section{Introduction}\label{sec01}

Electromagnetic fields can couple to gravity beyond the minimal gauge coupling.
In a curved spacetime,
one-loop vacuum polarization induces an effective photon action that links the field strength directly to the curvature tensor~\cite{ref01},
and a variety of effective field theories and modified gravity models produce the same structure~\cite{ref02, ref03, ref04}.
One representative form couples the field to the Weyl tensor,
$\alpha\,\Cwey{}^{\mu\nu\rho\sigma}F_{\mu\nu}F_{\rho\sigma}$,
with a single coupling constant $\alpha$ of dimension length squared.
The coupling acts differently on the two photon polarization states,
so the vacuum around a compact object becomes a birefringent medium.
Following Refs.~\cite{ref05, ref06} we label the two states PPL and PPM: polarization along the $l^{\mu}$ direction (in the plane of motion) and along $m^{\mu}$ (perpendicular to it);
on the Schwarzschild background they coincide with the even- and odd-parity sectors, respectively.
The two states are governed by different optical metrics,
show different absorption and scattering,
and cast different shadows.
This curvature-induced birefringence is distinct from the vacuum birefringence of strong-field quantum electrodynamics~\cite{ref07},
which requires an external electromagnetic field;
here the background curvature itself splits the polarizations.

The magnitude of every effect below is set by the dimensionless ratio $\alpha/M^2$ between the coupling and the horizon curvature scale;
Sec.~\ref{sec02} discusses its physically expected size and how the values we display should be read.
Two observational windows probe it.
Geometric optics (shadows, photon rings, ray deflection) probes the coupled photon-sphere structure and dominates horizon-scale images.
Wave optics (absorption, differential cross sections, the glory) responds through the barrier transmission and the full energy dependence of the phase,
and it keeps a polarization sensitivity that time-averaged images lose.
Combining the two into a single constraint means treating both polarizations, in both regimes, on the same footing.

Which black holes could show these effects at a measurable level is a question of scales.
Because $\alpha$ is a constant of Nature with the dimension of length squared,
its effects grow with the local curvature,
and laboratory bounds on anomalous vacuum birefringence~\cite{ref49} already exclude any observable signature for the supermassive black holes imaged at event-horizon scale~\cite{ref08, ref09, ref10},
as quantified in Sec.~\ref{sec02}.
An order-unity $\alpha/M^2$ would instead require a black hole of primordial mass~\cite{ref50},
light enough for the QED coefficient alone to reach that value.
Such an object is far too small for its shadow to be resolved, so any measurable signature would sit in the wave-optical observables---absorption spectra, differential cross sections, the glory---while the geometric optics enters as the classical limit that anchors them rather than as an imaging target.
The two respond to $\alpha$ through different combinations of the field equations, so we carry both through the same calculation.

Previous studies have explored various aspects of this scenario.
Research into the scattering and absorption of massless fields by black holes is now well established~\cite{ref11, ref12, ref13, ref14, ref15, ref16, ref17, ref18, ref19, ref20, ref21}.
The fundamental partial wave method has been implemented in a variety of contexts,
including regular and non-commutative black holes in higher dimensions.
In these cases, the absorption cross section oscillates around the geometric optical trap region at high frequencies,
while the differential cross section demonstrates the strongest parameter dependence at large angles~\cite{ref22, ref23};
the same machinery is what we extend to the two Weyl polarizations here.
The master equation governing electromagnetic perturbations with a Weyl correction,
and its separation into two parities, was derived by Chen and Jing~\cite{ref05},
and the geometric-optics consequences (lensing, the double shadow, polarized images) have been studied in a series of works~\cite{ref06, ref24, ref25, ref26}.
What has been missing is a single numerically controlled treatment that carries both polarization states through the whole chain, from the partial-wave scattering matrix to the shadow and the photon ring.

This paper provides that treatment.
The paper is organized as follows.
Section~\ref{sec02} introduces the Weyl-coupled photon dynamics and the separation into two parities on a Schwarzschild background.
Section~\ref{sec03} sets out the scattering matrix and the polarization-resolved wave observables it yields.
Section~\ref{sec04} constructs the absorption and greybody factors together with the shadow,
and Section~\ref{sec05} the differential cross section and a backward birefringence observable that vanishes identically at $\alpha=0$.
Section~\ref{sec06} computes the strong-field geometric optics, the photon ring and null geodesics,
and Section~\ref{sec07} discusses the observational implications and the regime in which they apply.
We use geometrized units $G=c=1$, signature $(-,+,+,+)$, mass $M$, and coupling $\alpha$.

\begin{figure}[t]
  \centering\includegraphics[width=\textwidth]{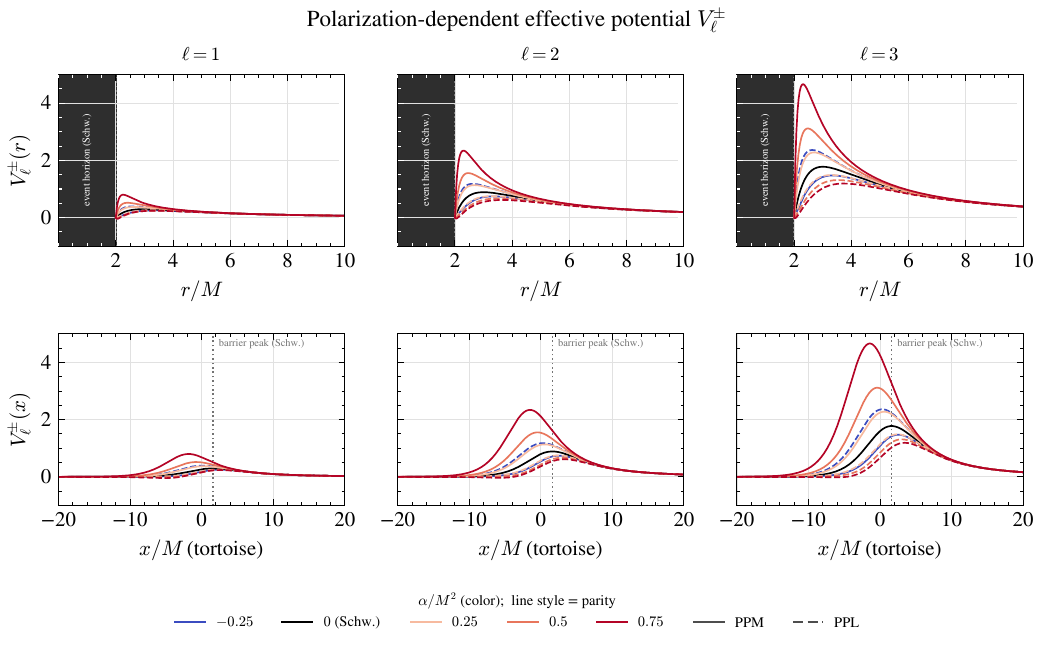}
  \caption{
  Effective potential of Eq.~\eqref{eq04} for the two polarizations (PPM solid, PPL dashed) and five values of $\alpha/M^2$ (color; the Schwarzschild case $\alpha=0$ in black).
  Top row:
  versus the areal radius $r$ for $\ell=1,2,3$, with the shaded band marking the black-hole interior $r\le2M$ and the dotted line the event horizon.
  Bottom row:
  versus the tortoise coordinate $x/M$ for the same $\ell$, with the dotted line at the Schwarzschild barrier peak.
  The barrier height moves in opposite directions for the two parities.}
  \label{fig01}
\end{figure}

\begin{figure}[t]
  \centering\includegraphics[width=\textwidth]{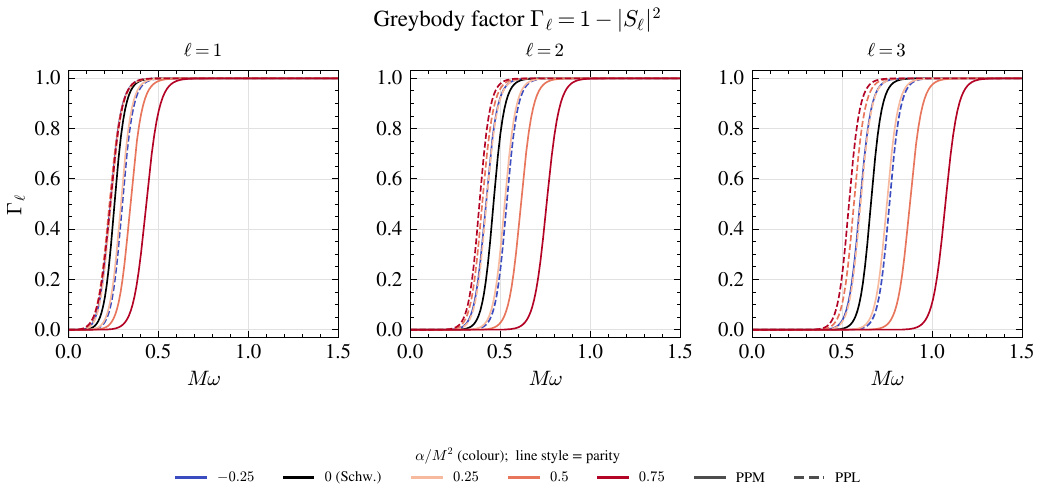}
  \caption{
  For the three lowest-order partial waves $\ell=1,2,3$,
  a plot of the greybody factor $\Gamma^{\pm}_\ell=1-|S^{\pm}_\ell|^2$ versus $M\omega$
  (PPM: solid line; PPL: dashed line; color determined by $\alpha/M^2$; Schwarzschild curve shown in black).
  Each partial wave turns on in a sigmoid step centered at $M\omega_{1/2}\simeq(\ell+\tfrac12)/(\bc/M)$, shifting outward as $\ell$ increases;
  the coupling splits each threshold---PPM (contracting $\bc$) turns on later,
  PPL (expanding $\bc$) earlier---the modulus-side counterpart of the polarization phase splitting.}
  \label{fig02}
\end{figure}

\begin{figure}[t]
  \centering\includegraphics[width=\textwidth]{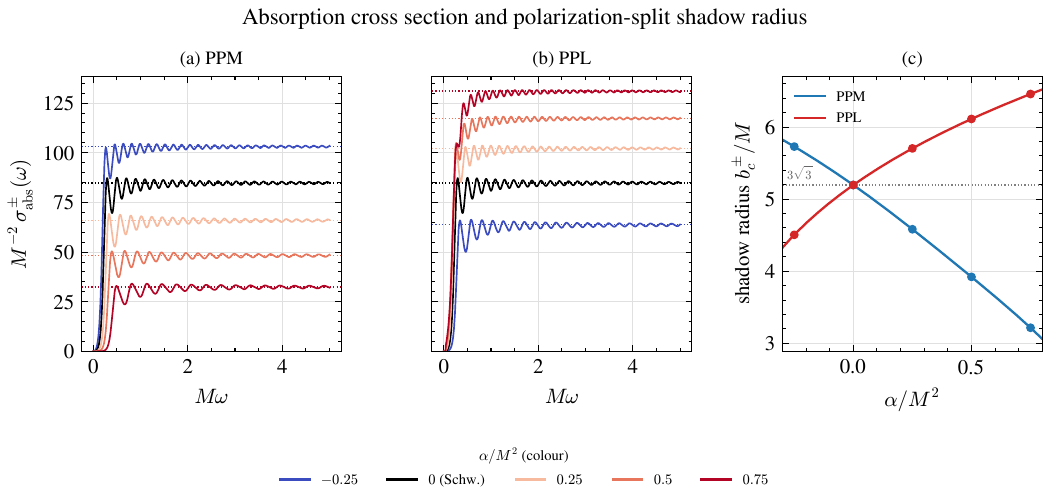}
  \caption{Absorption cross section $M^{-2}\sigma^{\pm}_{\mathrm{abs}}(\omega)$ for \textbf{(a)} PPM and \textbf{(b)} PPL and several $\alpha$;
  the color-matched dotted line on each curve marks its geometric-optics capture area $\pi(\bc/M)^2$, the high-frequency limit it oscillates toward.
  \textbf{(c)} The polarization-split shadow radius (critical impact parameter) $\bc^{\pm}/M$ versus $\alpha/M^2$, from the geometric-optics potential of Eq.~\eqref{eq07}:
  the two parity modes are degenerate at the Schwarzschild value $3\sqrt{3}\,M$ (dotted line) and separate as $|\alpha|$ grows, reaching $\bc^{\mathrm{PPM}}/M=3.22$ and $\bc^{\mathrm{PPL}}/M=6.46$ ($62\%$ apart) at $\alpha/M^2=0.75$.}
  \label{fig03}
\end{figure}

\section{Weyl-coupled photon dynamics}\label{sec02}

\subsection{Action and field equations}
The model is defined by the action of the electromagnetic field on a fixed gravitational background,
\begin{equation}
  S = \int \dd^4x\,\sqrt{-g}\left[
      -\tfrac{1}{4}F_{\mu\nu}F^{\mu\nu}
      + \alpha\, \Cwey{}^{\mu\nu\rho\sigma} F_{\mu\nu}F_{\rho\sigma}\right],
  \label{eq01}
\end{equation}
whose variation with respect to $A_\mu$ gives the modified Maxwell equation
\begin{equation}
  \nabla_\mu\!\left(F^{\mu\nu}
    - 4\alpha\,\Cwey{}^{\mu\nu\rho\sigma}F_{\rho\sigma}\right)=0 .
  \label{eq02}
\end{equation}
The photon is thus a test field:
the backreaction of the electromagnetic field on the spacetime metric is neglected throughout,
and the background is the fixed vacuum geometry specified below.
The Weyl tensor in Eq.~\eqref{eq02} contracts the field strength with the background curvature;
in vacuum it equals the Riemann tensor,
so the correction couples the photon directly to the tidal field that deflects light rays.
The coupling $\alpha$ has the dimension of length squared;
in physical terms it is a low-energy remnant of one-loop quantum-electrodynamic vacuum polarization in a curved background~\cite{ref01},
and the same operator arises generically in effective field theories of photon--gravity interactions~\cite{ref27}.
We adopt the normalization of Chen and Jing~\cite{ref05}.

The physically expected size of $\alpha$ deserves an honest statement.
For the one-loop QED (Drummond--Hathrell) origin the coefficient is set by the electron Compton wavelength,
$\alpha\sim(\alpha_{\mathrm{em}}/4\pi)\,\lambda_e^{2}$ with $\lambda_e$ the reduced electron Compton wavelength~\cite{ref01, ref27},
so $\alpha/M^2\sim10^{-4}(\lambda_e/M)^2$ is of order $10^{-35}$ for a solar-mass black hole and $10^{-55}$ for M87$^*$, both far too small to observe.
A larger $\alpha$ would not change where the coupling is best tested.
Since $\alpha$ takes the same value everywhere, the size of its effect at a given place is set by the curvature there;
whether that effect can be seen also depends on how precisely it can be measured.
The curvature enters through the Kretschmann scalar $K=R_{\mu\nu\rho\sigma}R^{\mu\nu\rho\sigma}$,
which equals $48M^2/r^6$ on a Schwarzschild background and, in vacuum, equals $\Cwey_{\mu\nu\rho\sigma}\Cwey^{\mu\nu\rho\sigma}$, the square of the tensor that $\alpha$ multiplies.
$K$ is about $10^{-32}\,\mathrm{km}^{-4}$ at the Earth's surface and $10^{-33}\,\mathrm{km}^{-4}$ at the Sun's,
but below $10^{-40}\,\mathrm{km}^{-4}$ at the horizon of M87$^*$, eight orders of magnitude below the terrestrial value.
Laboratory searches for anomalous vacuum birefringence therefore bound $\alpha$ directly:
taking the Weyl curvature at the Earth's surface, $\sim3\times10^{-23}\,\mathrm{m}^{-2}$,
and a PVLAS-class sensitivity to a polarization-dependent refractive index, $\Delta n\sim10^{-22}$~\cite{ref49},
gives the order-of-magnitude bound $\alpha\lesssim$ a few $\mathrm{m}^2$,
which translates into $\alpha/M^2\lesssim10^{-26}$ for M87$^*$.
Curvature alone does not decide the comparison for every black hole.
At the horizon of Sgr~A$^*$ the Kretschmann scalar reaches $\sim10^{-28}\,\mathrm{km}^{-4}$,
four orders of magnitude above the terrestrial value, and $\alpha/M^2\lesssim10^{-19}$ there.
The laboratory still wins on precision:
Event Horizon Telescope imaging pins the shadow diameter to about $10\%$~\cite{ref46, ref47, ref48},
whereas the searches quoted above work at $\Delta n\sim10^{-22}$,
so some twenty orders of magnitude in measurement precision outweigh four in curvature.
Observable effects on the shadow or other lensing features of supermassive (or stellar-mass) black holes are therefore excluded for the foreseeable future.
Order-unity couplings would instead require a far lighter black hole---a primordial one~\cite{ref50}.
Since $\alpha$ is fixed, $\alpha/M^2$ grows as $M$ falls:
the same QED coefficient gives $\alpha/M^2=1$ at $M\simeq5\times10^{15}\,\mathrm{g}$,
and hence $\alpha/M^2\gtrsim1$ for $M\lesssim10^{15}\,\mathrm{g}$,
with $\alpha$ itself at its Standard-Model value and every laboratory bound respected.
This window carries its own caveats:
objects below $\simeq5\times10^{14}\,\mathrm{g}$ have evaporated by now,
and the abundance of somewhat heavier ones is bounded by the diffuse gamma-ray background~\cite{ref50}.
That makes it the one mass range the laboratory bounds do not already close off,
not one where such couplings are predicted.
Throughout this paper we accordingly treat $\alpha$ as a phenomenological parameter of the effective theory,
and the values we display, up to $\alpha/M^2=0.75$, as a range that resolves the polarization splitting clearly and that could be physically motivated only for black holes of primordial mass;
the observational discussion of Sec.~\ref{sec07} is framed accordingly.

Two conditions bound the admissible couplings.
Regularity of the effective potentials outside the horizon requires
$\alpha<M^2$ for odd parity and $-M^2/2<\alpha<M^2$ for even parity~\cite{ref05}.
Dynamical stability is more restrictive for the odd sector:
the time-domain evolutions of Ref.~\cite{ref05} show that odd-parity perturbations grow exponentially for $\alpha$ below a negative, $\ell$-dependent threshold $\alpha_c(\ell)$,
which rises from $\simeq-1.15\,M^2$ at $\ell=1$ toward $-M^2/2$ as $\ell\to\infty$,
while the even sector is stable throughout its allowed range.
The common safe window is therefore $-M^2/2<\alpha<M^2$,
and the values used below, $\alpha/M^2\in\{0,\,-\tfrac14,\,+\tfrac14,\,+\tfrac12,\,+\tfrac34\}$,
lie inside it for every $\ell$.

\subsection{Background and master equation}
We work on the Schwarzschild metric with $f(r)=1-2M/r$.
Following Chen and Jing~\cite{ref05},
the perturbation $A_\mu$ is expanded in vector spherical harmonics of definite parity (odd,
transforming as $(-1)^{\ell+1}$,
and even, as $(-1)^{\ell}$), which splits Eq.~\eqref{eq02} into two decoupled sectors.
Eliminating the redundant components and separating the time and angular dependence reduces each parity to a single Schrödinger-type radial equation for
$\psi^{\pm}_{\ell}$~\cite{ref05},
\begin{equation}
  \frac{\dd^2\psi^{\pm}_{\ell}}{\dd \rs^2}
   +\left[\omega^2-\Veff(r)\right]\psi^{\pm}_{\ell}=0,
  \qquad \frac{\dd \rs}{\dd r}=\frac{1}{f(r)},
  \label{eq03}
\end{equation}
with the polarization-dependent effective potential
\begin{equation}
  \Veff(r)= f(r)\left[\frac{\ell(\ell+1)}{r^2}\,\mathcal{W}^{\pm}(r)
            +\mathcal{C}^{\pm}(r)\right].
  \label{eq04}
\end{equation}
The leading ``light-cone'' factor~\cite{ref01, ref02},
\begin{equation}
  \mathcal{W}^{\mathrm{PPM}}=\frac{r^3+16\alpha M}{r^3-8\alpha M},
  \qquad
  \mathcal{W}^{\mathrm{PPL}}=\frac{r^3-8\alpha M}{r^3+16\alpha M},
  \label{eq05}
\end{equation}
sets the geometric-optics (eikonal) dynamics and the photon-sphere splitting,
while the sub-leading curvature term,
\begin{align}
  \mathcal{C}^{\mathrm{PPM}} &= -\frac{24\alpha M\,(2r^4-5Mr^3-10\alpha M r
       +28\alpha M^2)}{r^3\,(r^3-8\alpha M)^2},\nonumber\\
  \mathcal{C}^{\mathrm{PPL}} &= +\frac{24\alpha M\,(2r^4-5Mr^3+2\alpha M r
       +4\alpha M^2)}{r^3\,(r^3-8\alpha M)^2},
  \label{eq06}
\end{align}
is an $O(\alpha)$ piece that we keep in the integration;
it leaves the photon-sphere structure unchanged but contributes to the wave-optics phase shifts.
(Both parities share the denominator $r^3(r^3-8\alpha M)^2$, as in Eqs.~(16) and (17) of Ref.~\cite{ref05};
for the couplings admitted below its zero lies inside the horizon.)
PPL (even parity) and PPM (odd parity) denote the two photon polarizations.
At $\alpha=0$ both factors reduce to unity,
the curvature term vanishes,
and Eq.~\eqref{eq04} becomes the minimal Maxwell potential,
with the polarizations degenerate and no birefringence.
For $\alpha\neq0$ the two potentials differ,
and since $\mathcal{W}^{\mathrm{PPM}}=1/\mathcal{W}^{\mathrm{PPL}}$ the leading factors move in opposite directions, which is the source of every polarization-dependent effect below.
Figure~\ref{fig01} shows the resulting barriers.
The two potentials can be cast in the superpartner-like form
$V^{\mathrm{odd,even}}=W_s^2\pm\dd W_s/\dd\rs+\beta$ of Ref.~\cite{ref05};
because the remainder $\beta(r)$ is a function of $r$ rather than a constant factorization energy,
this relation does not make the pair isospectral,
and the transmission moduli of the two parities indeed differ (Fig.~\ref{fig02}).
What the reciprocity $\mathcal{W}^{\mathrm{PPM}}=1/\mathcal{W}^{\mathrm{PPL}}$ does enforce is a splitting that is nearly antisymmetric under $(\mathrm{parity},\alpha)\to(\overline{\mathrm{parity}},-\alpha)$ at leading order.
All backgrounds used here lie inside the stability window stated above,
so the scattering matrix is well defined for every $(\omega,\ell)$.

\subsection{Geometric-optics limit}
In this limit Eq.~\eqref{eq02} reduces to null propagation on two polarization-dependent optical metrics:
the angular sector is rescaled by the light-cone factor $\mathcal{W}^{\pm}$,
so each polarization sees the effective radial potential
\begin{equation}
  V^{\pm}_{\mathrm{geo}}(r)=\frac{f(r)}{r^2}\,\mathcal{W}^{\pm}(r),
  \label{eq07}
\end{equation}
with its own photon sphere $r_{\mathrm{ph}}^{\pm}(\alpha)$,
given by $\dd V^{\pm}_{\mathrm{geo}}/\dd r=0$,
and its own critical impact parameter
$\bc^{\pm}=[V^{\pm}_{\mathrm{geo}}(r_{\mathrm{ph}}^{\pm})]^{-1/2}$~\cite{ref25, ref24}.
The circular photon orbit is unstable.
At $\alpha=0$ these collapse to the Schwarzschild values $r_{\mathrm{ph}}=3M$ and $\bc=3\sqrt{3}\,M\simeq5.196\,M$.
The splitting grows with $\alpha$ and,
to leading order,
is symmetric between the parities through $\mathcal{W}^{\mathrm{PPM}}=1/\mathcal{W}^{\mathrm{PPL}}$:
at $\alpha/M^2=0.5$ the PPM photon sphere contracts to $r_{\mathrm{ph}}=2.49\,M$ with $\bc=3.92\,M$,
while PPL expands to $r_{\mathrm{ph}}=3.58\,M$, $\bc=6.11\,M$,
so the two polarizations bound visibly different shadows.
Table~\ref{tab01} collects $r_{\mathrm{ph}}^{\pm}$, $\bc^{\pm}$,
and the photon-ring Lyapunov exponents $\gamma^{\pm}$ (Sec.~\ref{sec06}) for all couplings used in this work.

\begin{table}[t]
  \caption{Geometric-optics landscape of the two polarizations versus the coupling:
  photon-sphere radius $r_{\mathrm{ph}}^{\pm}$, critical impact parameter $\bc^{\pm}$,
  and Lyapunov exponent $\gamma^{\pm}$ of the unstable photon orbit in units of the Schwarzschild value
  ($\gamma$ is quoted per half-orbit, so successive photon subrings are demagnified by $e^{-\gamma}$;
  the Schwarzschild value is $\gamma=\pi$).}
  \label{tab01}
  \centering
  \begin{tabular*}{\textwidth}{@{\extracolsep{\fill}}r cc c cc c}
  \toprule
   & \multicolumn{3}{c}{PPM (odd)} & \multicolumn{3}{c}{PPL (even)}\\
  \cmidrule(lr){2-4}\cmidrule(lr){5-7}
  $\alpha/M^2$ & $r_{\mathrm{ph}}/M$ & $\bc/M$ & $\gamma/\pi$ & $r_{\mathrm{ph}}/M$ & $\bc/M$ & $\gamma/\pi$\\
  \midrule
  $-0.25$ & $3.344$ & $5.729$ & $0.954$ & $2.639$ & $4.505$ & $1.237$\\
  $0$     & $3$     & $5.196$ & $1$     & $3$     & $5.196$ & $1$\\
  $0.25$  & $2.704$ & $4.582$ & $1.219$ & $3.310$ & $5.703$ & $0.937$\\
  $0.5$   & $2.491$ & $3.921$ & $1.703$ & $3.578$ & $6.112$ & $0.921$\\
  $0.75$  & $2.320$ & $3.215$ & $2.649$ & $3.814$ & $6.458$ & $0.920$\\
  \bottomrule
  \end{tabular*}
\end{table}

\section{Scattering matrix and observables}\label{sec03}

\subsection{Boundary conditions and the scattering matrix}
We apply a purely ingoing wave at the horizon,
where $\rs\to-\infty$ and a plane wave is exact,
and an incoming-plus-outgoing combination of Riccati--Hankel functions at $\rs\to\infty$,
where the centrifugal tail of Eq.~\eqref{eq04} still survives,
\begin{equation}
  \psi^{\pm}_{\ell}\sim
  \begin{cases}
    e^{-i\omega \rs}, & \rs\to-\infty,\\[2pt]
    A^{\pm}_{\ell}\,\hat H^{-}_{\ell}(\omega\rs)+B^{\pm}_{\ell}\,\hat H^{+}_{\ell}(\omega\rs), & \rs\to+\infty,
  \end{cases}
  \label{eq08}
\end{equation}
with the Riccati--Hankel functions
$\hat H^{\pm}_{\ell}(z)=z\,[\,j_\ell(z)\pm i\,y_\ell(z)\,]$
the outgoing and ingoing solutions of the free centrifugal equation,
which carry the $1/\rs^2$ tail phase exactly.
The scattering-matrix element is then
$S^{\pm}_{\ell}=B^{\pm}_{\ell}/A^{\pm}_{\ell}=e^{2i\delta^{\pm}_{\ell}}$
for elastic modes (those not absorbed, $|S^{\pm}_{\ell}|=1$, so the wave only acquires a phase),
and $1-|S^{\pm}_{\ell}|^2$ is the absorption probability through the barrier of Eq.~\eqref{eq04}.
Every wave observable descends from this one object:
solving Eq.~\eqref{eq03} on an $(\omega,\ell)$ grid for $S^{\pm}_\ell$ gives the greybody factor and absorption from $|S_\ell|$,
the integrated scattering from $|1-S_\ell|$,
the phase shifts from $\arg S_\ell$,
and the differential cross section from the spin-1 amplitudes.
We read $S^{\pm}_\ell$ off by matching the Numerov solution to the Hankel basis of Eq.~\eqref{eq08} rather than to plane waves:
the surviving centrifugal tail would otherwise rob a plane-wave match of part of the accumulated phase and corrupt every phase-sensitive observable,
while the Hankel basis passes the $V\equiv0\Rightarrow S\equiv1$ self-test and reproduces the Schwarzschild glory at $163.4^\circ$~\cite{ref17}.

\subsection{Cross sections}
Each mode's greybody (transmission) factor is
$\Gamma^{\pm}_\ell=1-|S^{\pm}_\ell|^2$,
and the absorption and scattering cross sections are
\begin{equation}
  \sigma^{\pm}_{\mathrm{abs}}=\frac{\pi}{\omega^2}\sum_{\ell}(2\ell+1)\,
    \Gamma^{\pm}_\ell,\qquad
  \sigma^{\pm}_{\mathrm{sc}}=\frac{\pi}{\omega^2}\sum_{\ell}(2\ell+1)\,
    |1-S^{\pm}_\ell|^2 .
  \label{eq09}
\end{equation}
The absorption uses only the modulus $|S_\ell|$;
the differential cross section is the phase-sensitive observable.
For a spin-1 field the scattering amplitude separates into a helicity-preserving amplitude $f$ and a helicity-reversing amplitude $g$~\cite{ref17, ref18},
built from the parity combinations of the two scattering matrices,
\begin{equation}
  a^{+}_{\ell}=\tfrac12\!\left(S^{E}_{\ell}+S^{M}_{\ell}\right)-1,\qquad
  a^{-}_{\ell}=\tfrac12\!\left(S^{M}_{\ell}-S^{E}_{\ell}\right),
  \label{eq10a}
\end{equation}
where the $-1$ removes the forward plane-wave delta, as
\begin{align}
  f(\theta)&=\frac{1}{2i\omega}\sum_{\ell\ge1}(2\ell+1)\,a^{+}_{\ell}\,d^{\ell}_{1,1}(\theta),\nonumber\\
  g(\theta)&=\frac{1}{2i\omega}\sum_{\ell\ge1}(2\ell+1)\,a^{-}_{\ell}\,d^{\ell}_{1,-1}(\theta),
  \label{eq10b}
\end{align}
with the spin-1 Wigner functions
$d^{\ell}_{1,\pm1}=(\pi_\ell\pm\tau_\ell)/[\ell(\ell+1)]$,
$\pi_\ell(\cos\theta)=P^{1}_{\ell}/\sin\theta$ and $\tau_\ell=\dd P^{1}_{\ell}/\dd\theta$;
the unpolarized differential cross section is
\begin{equation}
  \frac{\dd\sigma}{\dd\Omega}=|f(\theta)|^2+|g(\theta)|^2 .
  \label{eq10c}
\end{equation}
The partial-wave series converge slowly because of the long-range gravitational tail;
we sum them with the Yennie reduction~\cite{ref28, ref17, ref29, ref21},
applying $m=2$ passes of the $(1-\cos\theta)$ coefficient recursion before evaluation (Sec.~\ref{sec03num}).
At the exact backward point $d^{\ell}_{1,1}(\pi)=0$ while $|d^{\ell}_{1,-1}(\pi)|=1$,
so the backward flux is $|g(\pi)|^2$ alone:
proportional to $|S^{M}-S^{E}|^2$,
it is a direct measure of birefringence that vanishes identically at $\alpha=0$.
For a single parity ($S^{E}=S^{M}$) the helicity-reversing amplitude vanishes, $g\equiv0$.
The forward $16M^2/\theta^4$ rise is the universal Newtonian divergence,
not a Weyl effect, and lies outside our scope.

\subsection{Numerical implementation and accuracy}\label{sec03num}
Equation~\eqref{eq03} is integrated by the Numerov method from a point just outside the horizon to the matching radius $\rs^{\max}=200\,M$,
with tortoise step $\Delta\rs=2\times10^{-3}\,M$,
where the solution is matched to the Riccati--Hankel pair of Eq.~\eqref{eq08};
the residual exterior phase from the potential tail beyond $\rs^{\max}$ is restored analytically through the $\omega$-independent integral
$K_\ell=\int_{\rs^{\max}}^{\infty}[V_\ell-\ell(\ell+1)/\rs^2]\,\dd\rs$, applied as $\Delta\delta_\ell=-K_\ell/(2\omega)$.
We carry $\ell\le200$ and accept a mode only within the matching-trust window $\omega\rs^{\max}\ge5(\ell+\tfrac12)$;
the amplitude sums of Eqs.~\eqref{eq10a}--\eqref{eq10c} use the Yennie reduction with $m=2$ and a cosine taper of the coefficient tail.
The frequency grids are $\Delta(M\omega)=5\times10^{-4}$ for the cross sections of Fig.~\ref{fig03} and $5\times10^{-3}$, up to $M\omega=8$, for the backward curve of Fig.~\ref{fig08b};
the latter is computed with the deeper settings $\ell\le400$ and $\rs^{\max}=300\,M$,
because the parity difference $a^{-}_{\ell}$ decays only as $1/\ell^{2}$ at high frequency.

The accuracy is controlled as follows.
(i)~\emph{Unitarity:} over the full production grid (five couplings, both parities, $1600$ frequencies, $\ell\le200$) the largest violation is $\max(|S^{\pm}_\ell|-1)=3.6\times10^{-10}$.
(ii)~\emph{Free-potential test:} setting $V\equiv0$ returns $|S_\ell-1|\le10^{-9}$ for all modes, so the matching itself introduces no spurious phase.
(iii)~\emph{Consistency of the angular sums:} the optical-theorem-type identity between $\sigma_{\mathrm{sc}}$ of Eq.~\eqref{eq09} and the angular integral of Eq.~\eqref{eq10c} holds to $4\times10^{-5}$.
(iv)~\emph{Convergence:} Table~\ref{tab02} varies the step, the matching radius, and $\ell_{\max}$ about the production values at the least favorable displayed point ($M\omega=2$, $\alpha/M^2=0.75$) and reports the change of the parity products $S^{M}_{\ell}\bar S^{E}_{\ell}$ and of the two most sensitive observables;
all are at or below the few-$\times10^{-3}$ level, far smaller than any effect discussed in this paper.
Repeating the sweep at $M\omega=8$ gives step and matching-radius entries at or below the same levels;
the $\ell$ truncation of the backward sum, probed by varying the summation window across the trusted range,
leaves a residual below $1\%$ for $M\omega\le5$ and ${\lesssim}10\%$ at the top of the displayed range, which is immaterial on the logarithmic scale of Fig.~\ref{fig08b} and does not shift the null positions.
(v)~\emph{Literature anchor:} at $\alpha=0$ the absorption spectra approach $\pi\bc^2$ (ratio $1.003$ at $M\omega=5$) and the glory peak at $M\omega=2$ lies at $164.0^\circ$, within $0.7^\circ$ of the $163.4^\circ$ of Crispino \emph{et al.}~\cite{ref17}.

\begin{table}[t]
  \caption{Convergence of the scattering matrix and of the two most phase-sensitive observables at $M\omega=2$, $\alpha/M^2=0.75$, relative to the production settings ($\Delta\rs=2\times10^{-3}M$, $\rs^{\max}=200M$, $\ell_{\max}=200$):
  maximum change of the parity products $S^{M}_{\ell}\bar S^{E}_{\ell}$ over the trusted modes,
  and relative changes of the exact-backward cross section $B=|g(\pi)|^2$ and of the glory height $G=\dd\sigma/\dd\Omega(163.4^\circ)$.}
  \label{tab02}
  \centering
  \begin{tabular*}{\textwidth}{@{\extracolsep{\fill}}l c c c}
  \toprule
  variation & $\max|\Delta(S^{M}\bar S^{E})|$ & $|\Delta B|/B$ & $|\Delta G|/G$\\
  \midrule
  $\Delta\rs/2$          & $2\times10^{-7}$ & $1\times10^{-6}$ & $2\times10^{-6}$\\
  $2\,\Delta\rs$         & $4\times10^{-7}$ & $3\times10^{-6}$ & $4\times10^{-6}$\\
  $\rs^{\max}=100M$      & $8\times10^{-4}$ & $2\times10^{-2}$ & $7\times10^{-3}$\\
  $\rs^{\max}=300M$      & $1\times10^{-4}$ & $2\times10^{-3}$ & $2\times10^{-3}$\\
  $\rs^{\max}=400M$      & $1\times10^{-4}$ & $1\times10^{-3}$ & $1\times10^{-3}$\\
  $\ell_{\max}=100$      & $0$ & $0$ & $0$\\
  $\ell_{\max}=300$      & $0$ & $0$ & $0$\\
  \bottomrule
  \end{tabular*}
\end{table}

\section{Greybody, absorption, and the shadow}\label{sec04}

\begin{figure}[t]
  \centering
  \includegraphics[width=0.48\columnwidth]{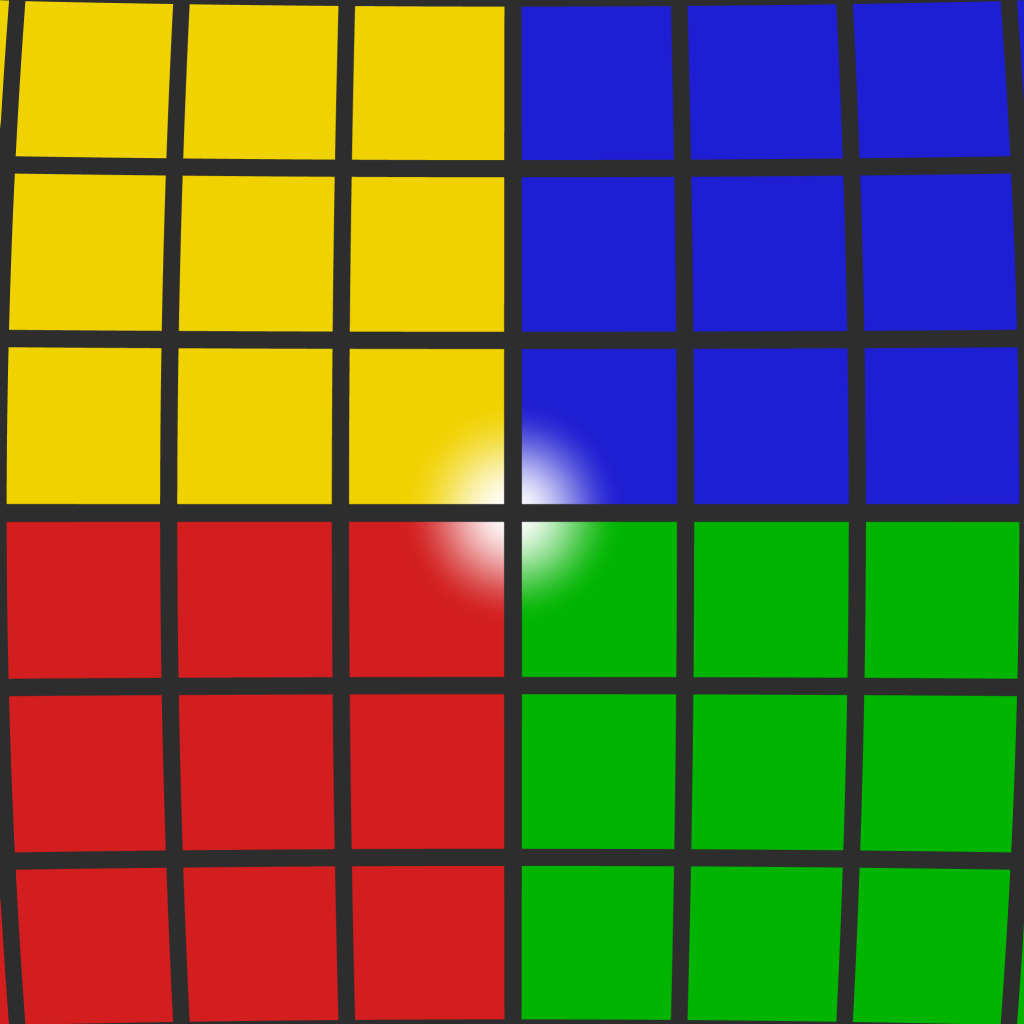}\hfill
  \includegraphics[width=0.48\columnwidth]{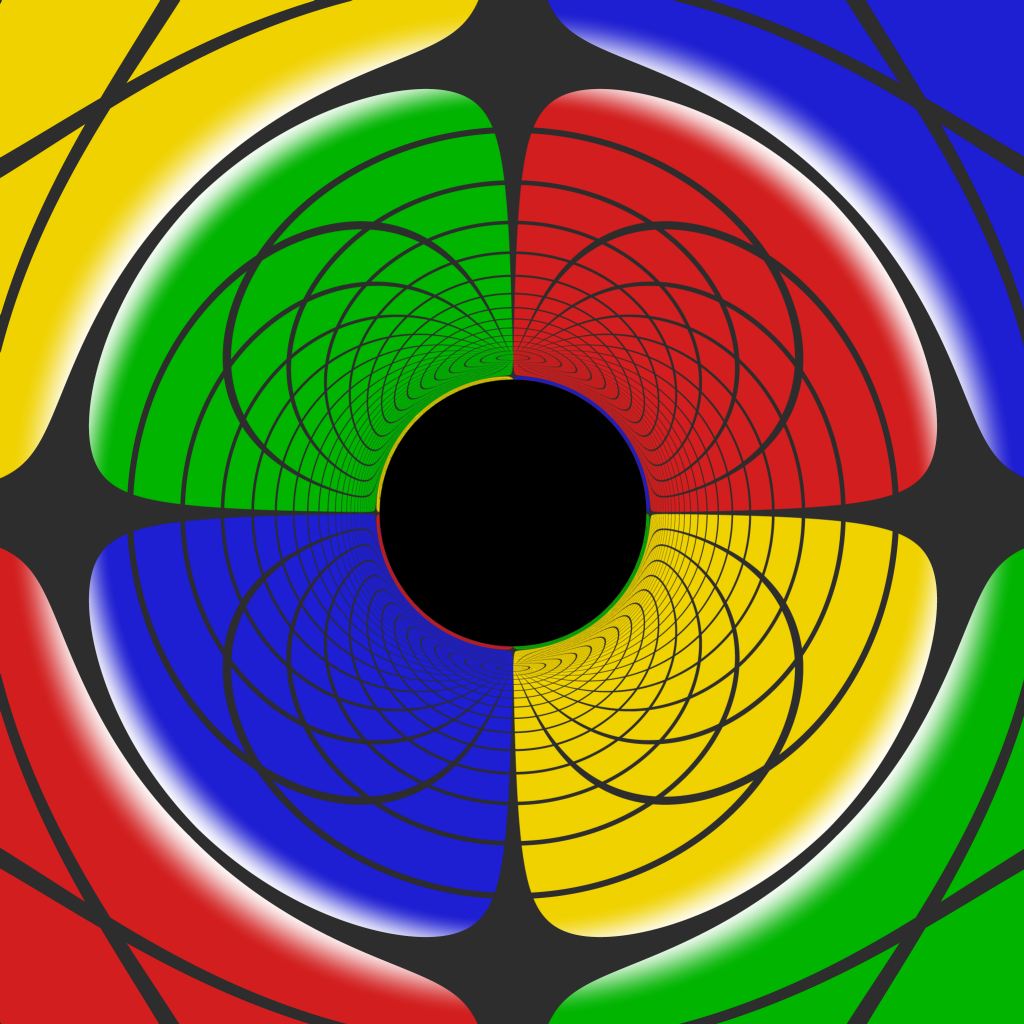}
  \caption{
  A four-color renderer in an orthogonal (impact parameter) camera.
  Left: a flat ($M=0$) reference image, in which the distributions in the four quadrants are analytical and there are no shadows.
  Right: a Schwarzschild diagram, showing the central shadow, the celestial sphere grid and the reference spot curved onto the Einstein ring.}
  \label{fig04}
\end{figure}

\begin{figure}[t]
  \centering
  \setlength{\tabcolsep}{4pt}\renewcommand{\arraystretch}{1.1}%
  \begin{tabular}{@{}c cccc@{}}
    & {\small$\alpha/M^2=-0.25$} & {\small$\alpha/M^2=0.25$}
      & {\small$\alpha/M^2=0.5$} & {\small$\alpha/M^2=0.75$}\\[3pt]
    \rotatebox[origin=c]{90}{\small PPM}
      & \shpanel{0.22\textwidth}{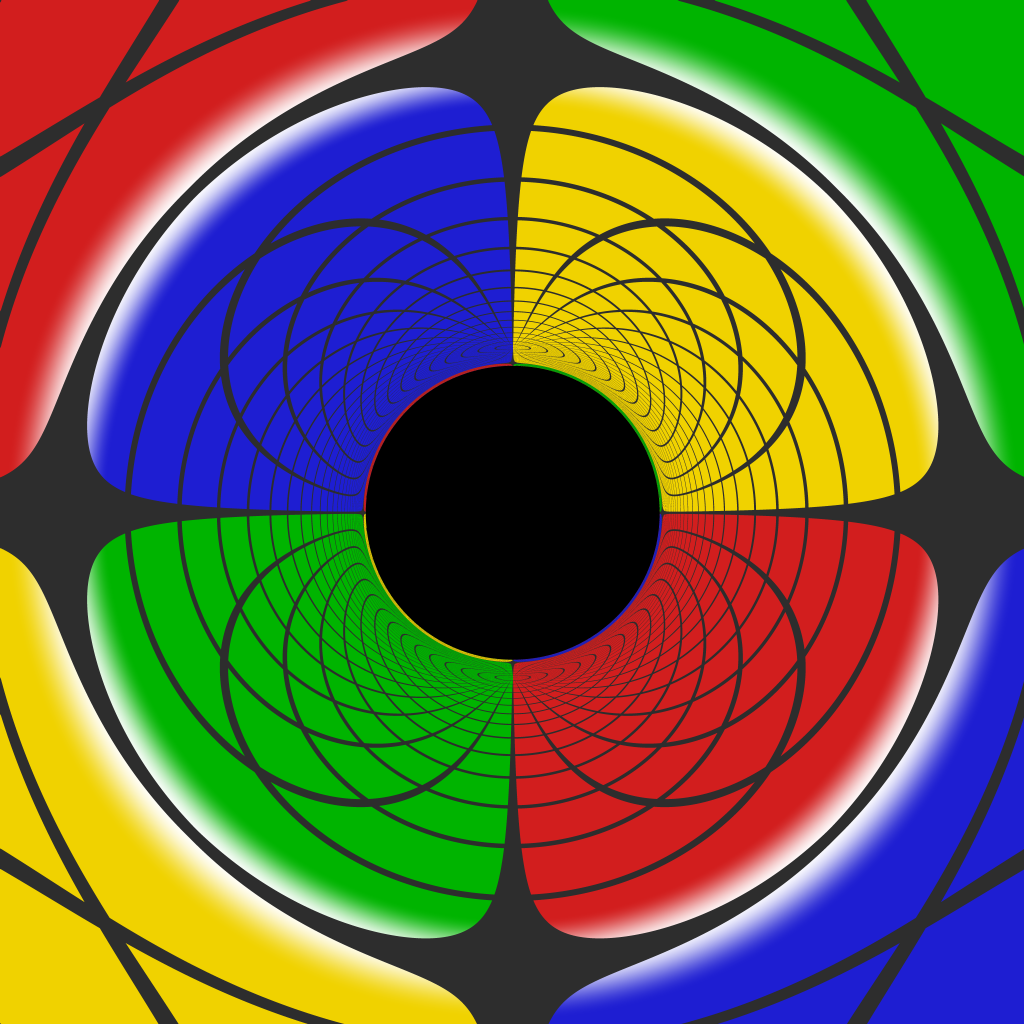}{5.73}
      & \shpanel{0.22\textwidth}{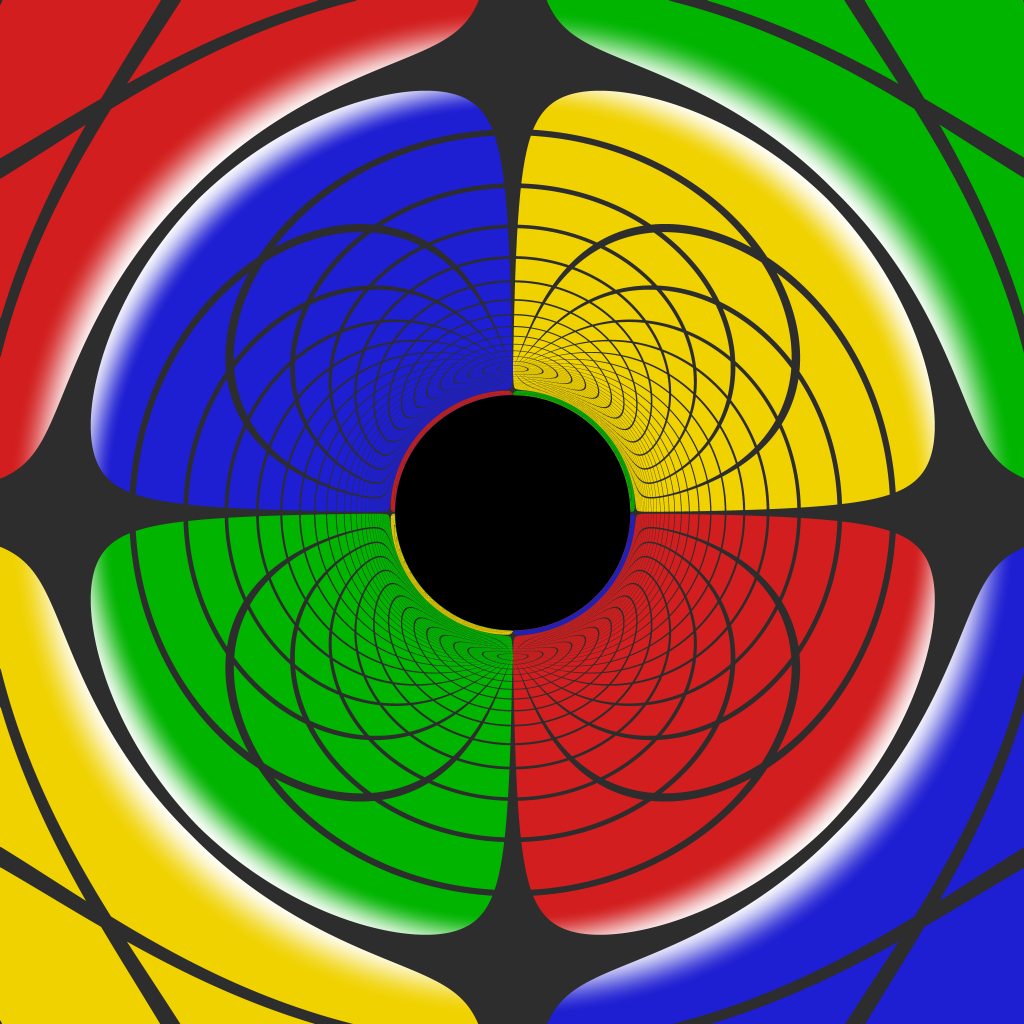}{4.58}
      & \shpanel{0.22\textwidth}{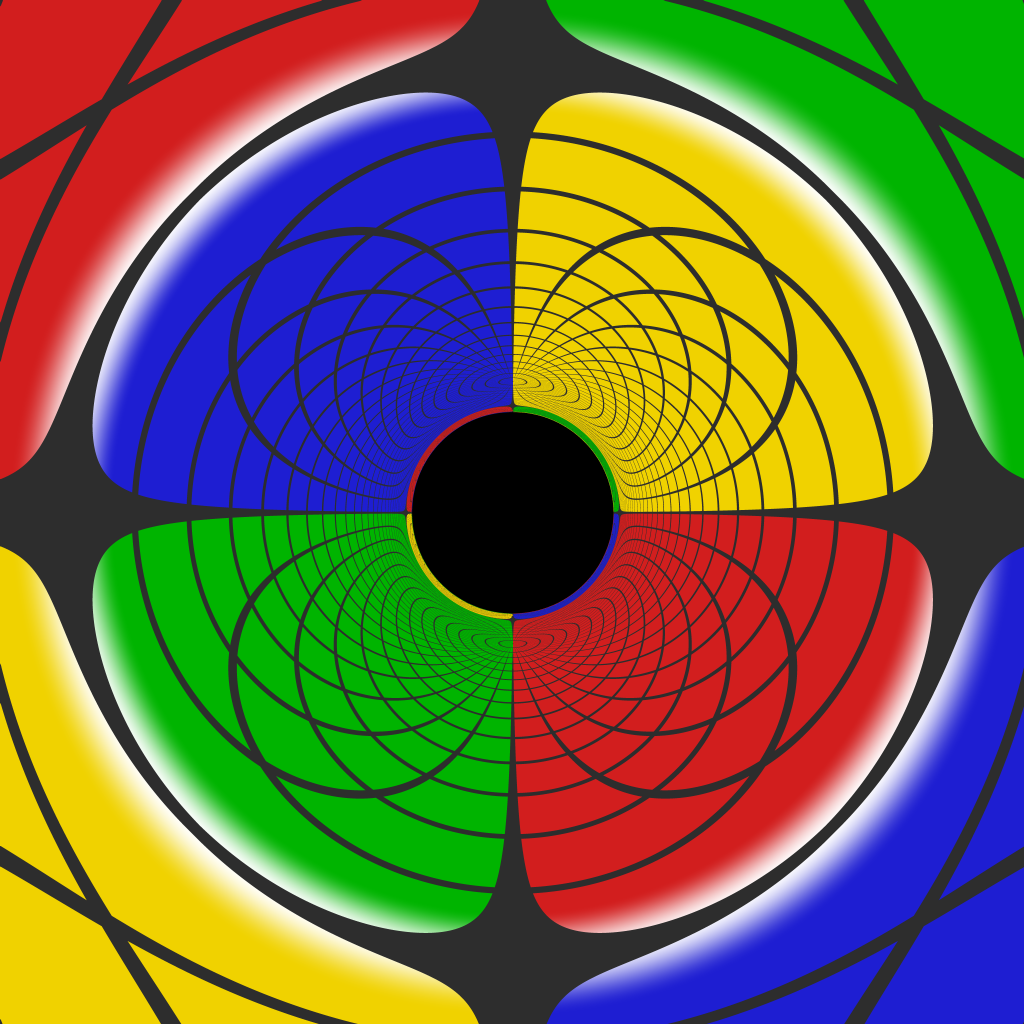}{3.92}
      & \shpanel{0.22\textwidth}{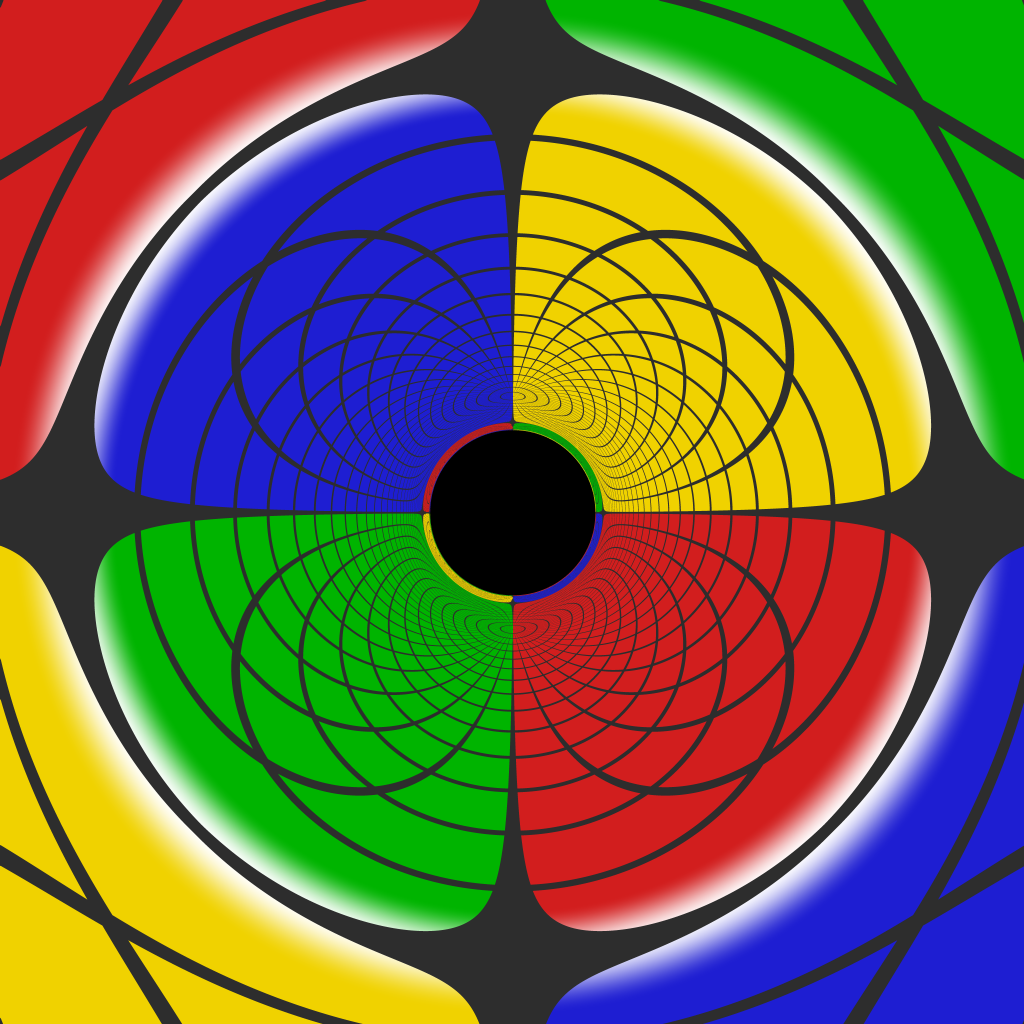}{3.22}\\[3pt]
    \rotatebox[origin=c]{90}{\small PPL}
      & \shpanel{0.22\textwidth}{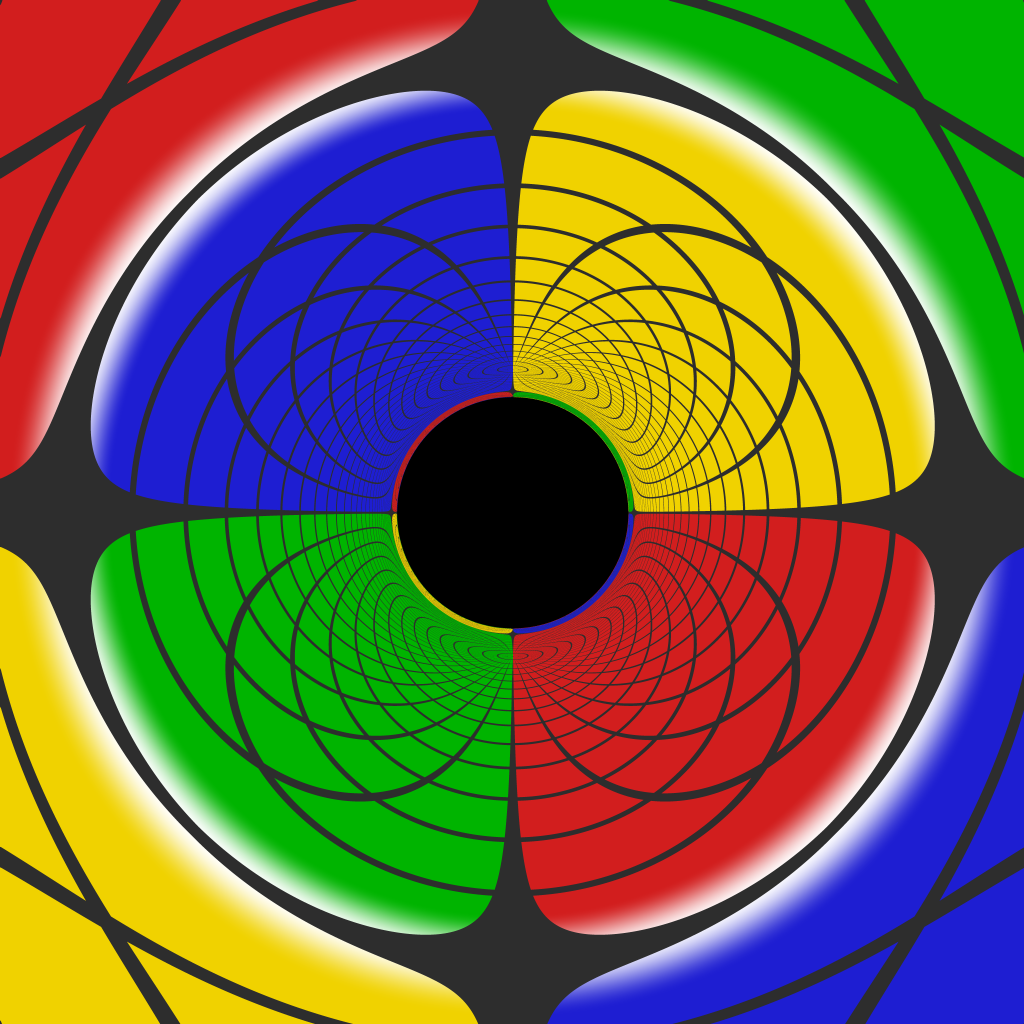}{4.51}
      & \shpanel{0.22\textwidth}{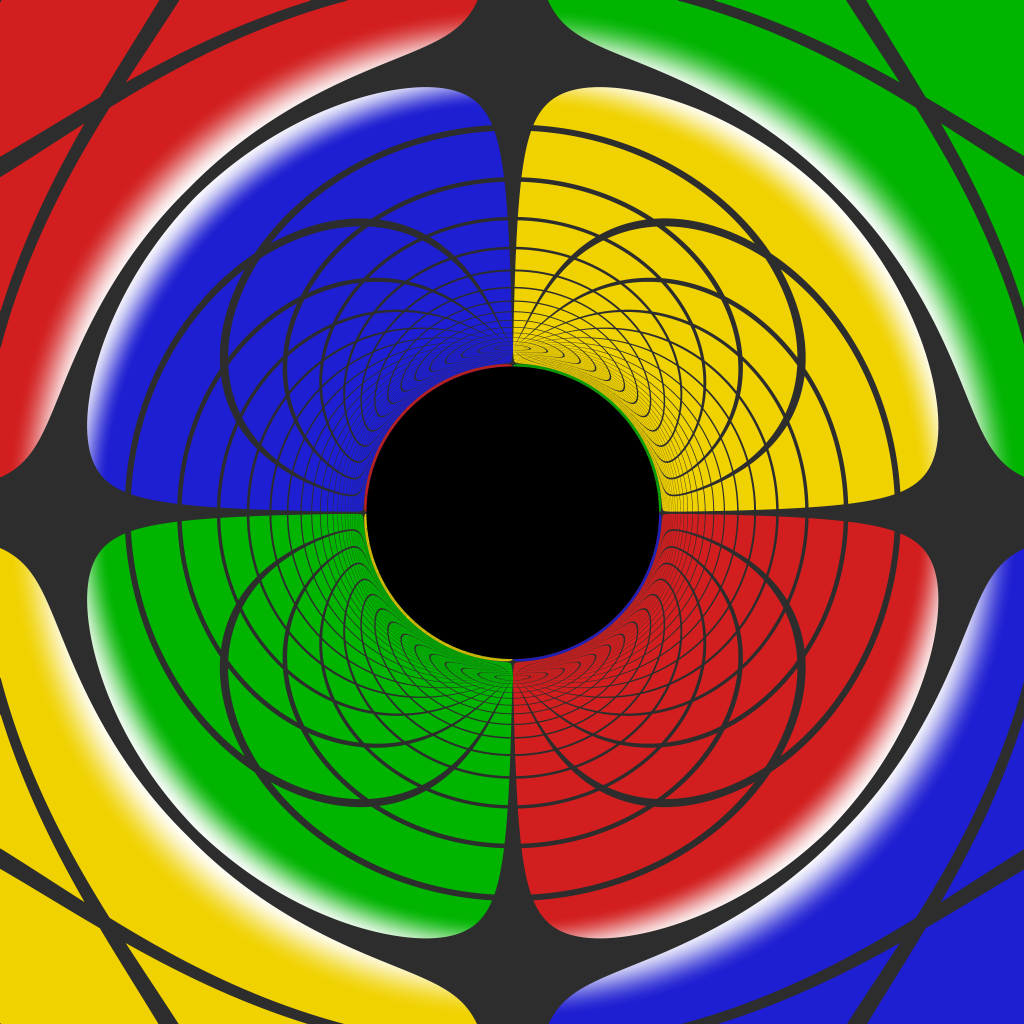}{5.70}
      & \shpanel{0.22\textwidth}{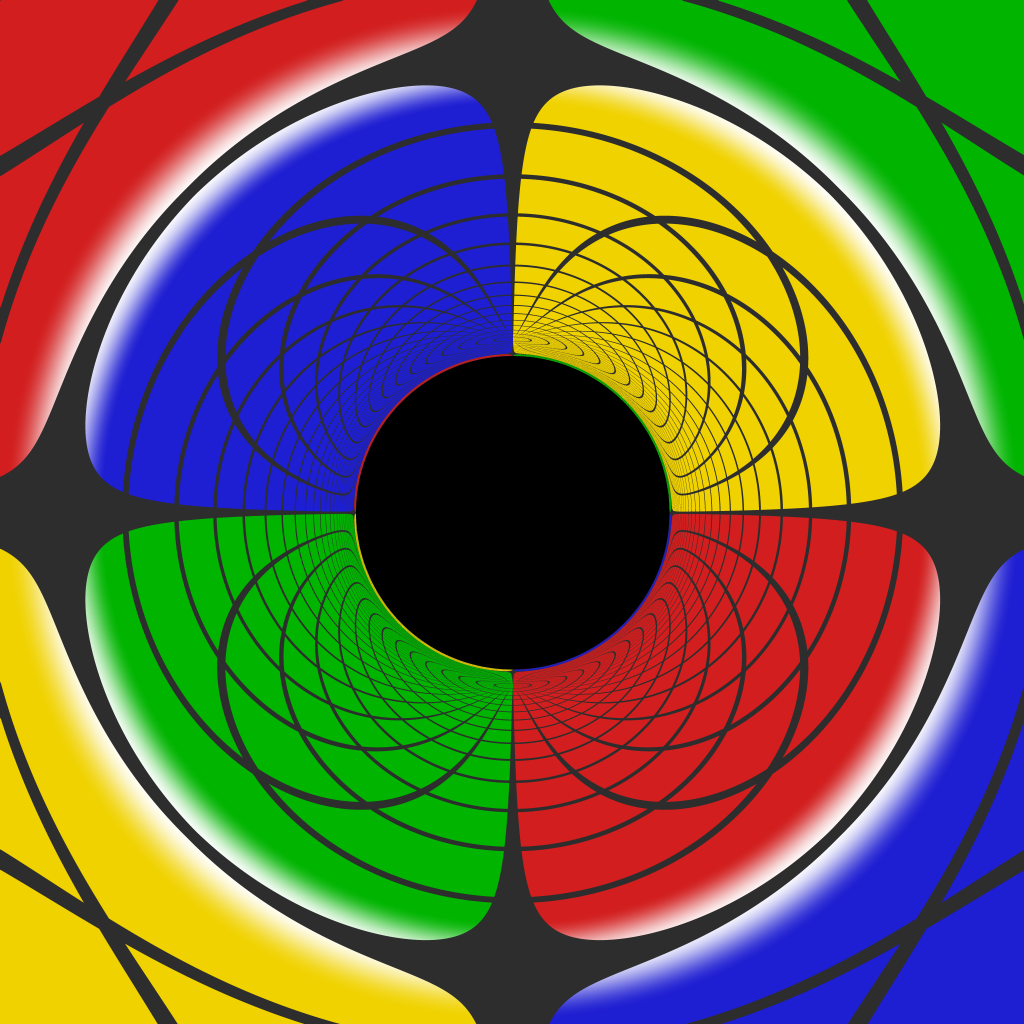}{6.11}
      & \shpanel{0.22\textwidth}{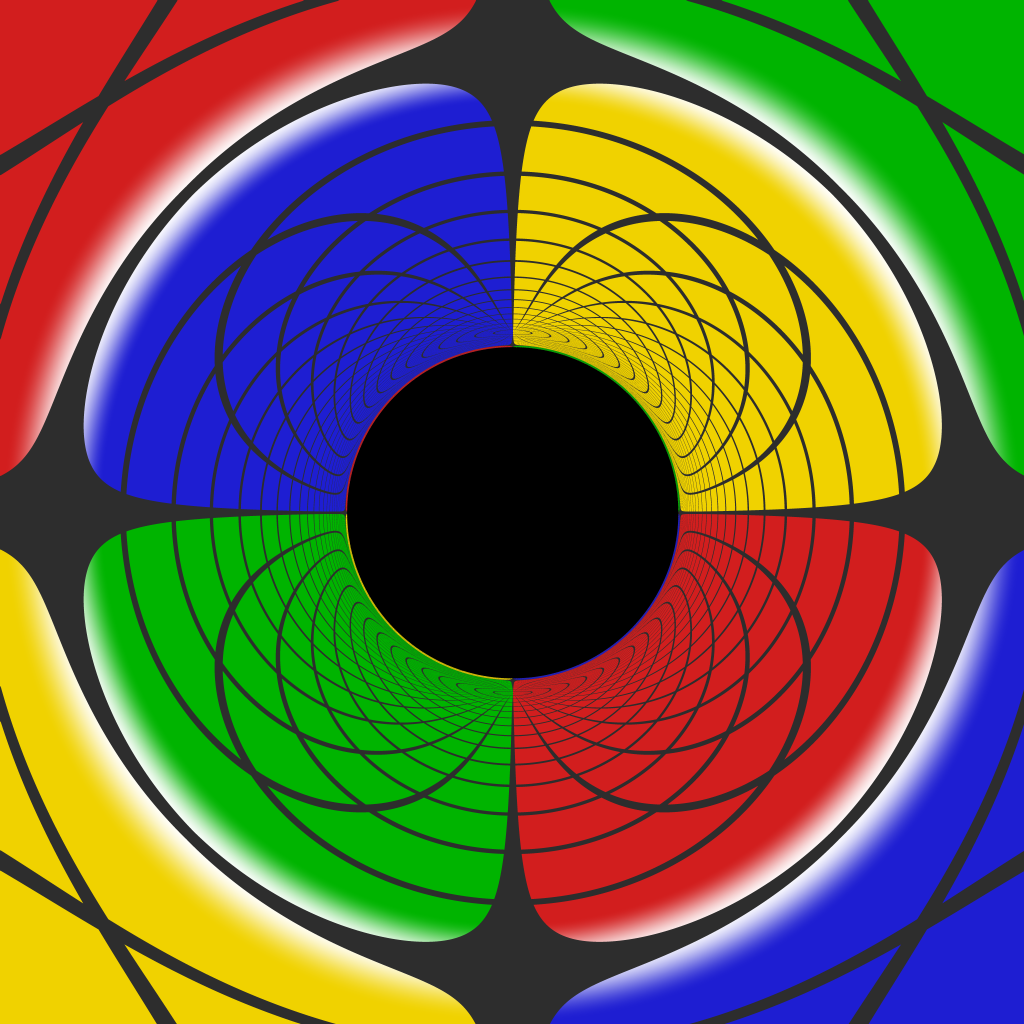}{6.46}\\
  \end{tabular}
  \caption{
  Polarization-dependent (double) shadow, as a grid over the two polarizations (PPM top, PPL bottom) and the four coupled cases $\alpha/M^2\in\{-0.25,0.25,0.5,0.75\}$ (left to right;
  the degenerate $\alpha=0$ case is omitted).
  The PPM edge contracts and the PPL edge expands as $\alpha$ grows, so the hole bounds two distinct critical curves.
  The critical impact parameter $\bc/M$ is printed below each panel
  (the omitted degenerate case $\alpha=0$ has the common value $\bc=3\sqrt3\,M\simeq5.196\,M$);
  the image plane spans $|b|\le20\,M$.}
  \label{fig05}
\end{figure}

\begin{figure}[t]
  \centering
  \setlength{\tabcolsep}{4pt}\renewcommand{\arraystretch}{1.05}%
  \begin{tabular}{@{}cccc@{}}
    \includegraphics[width=0.24\textwidth]{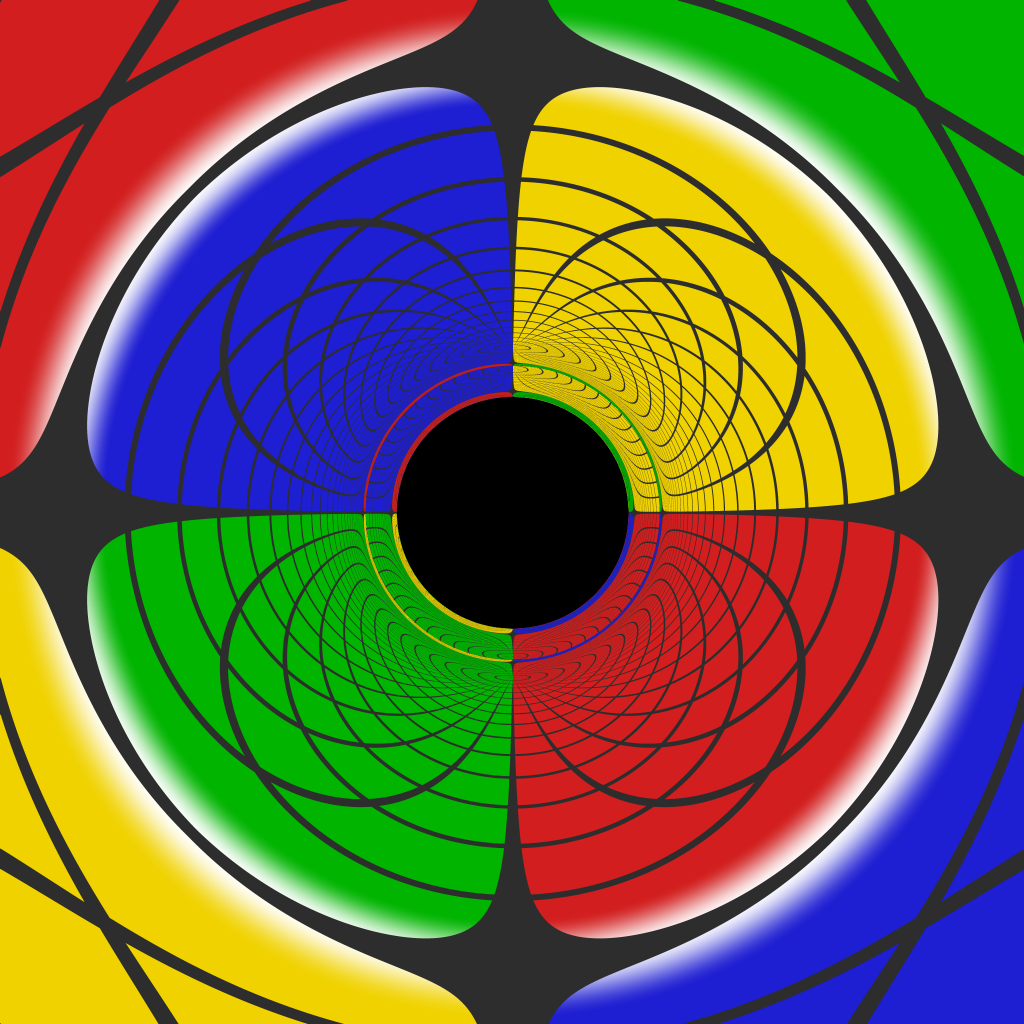} &
    \includegraphics[width=0.24\textwidth]{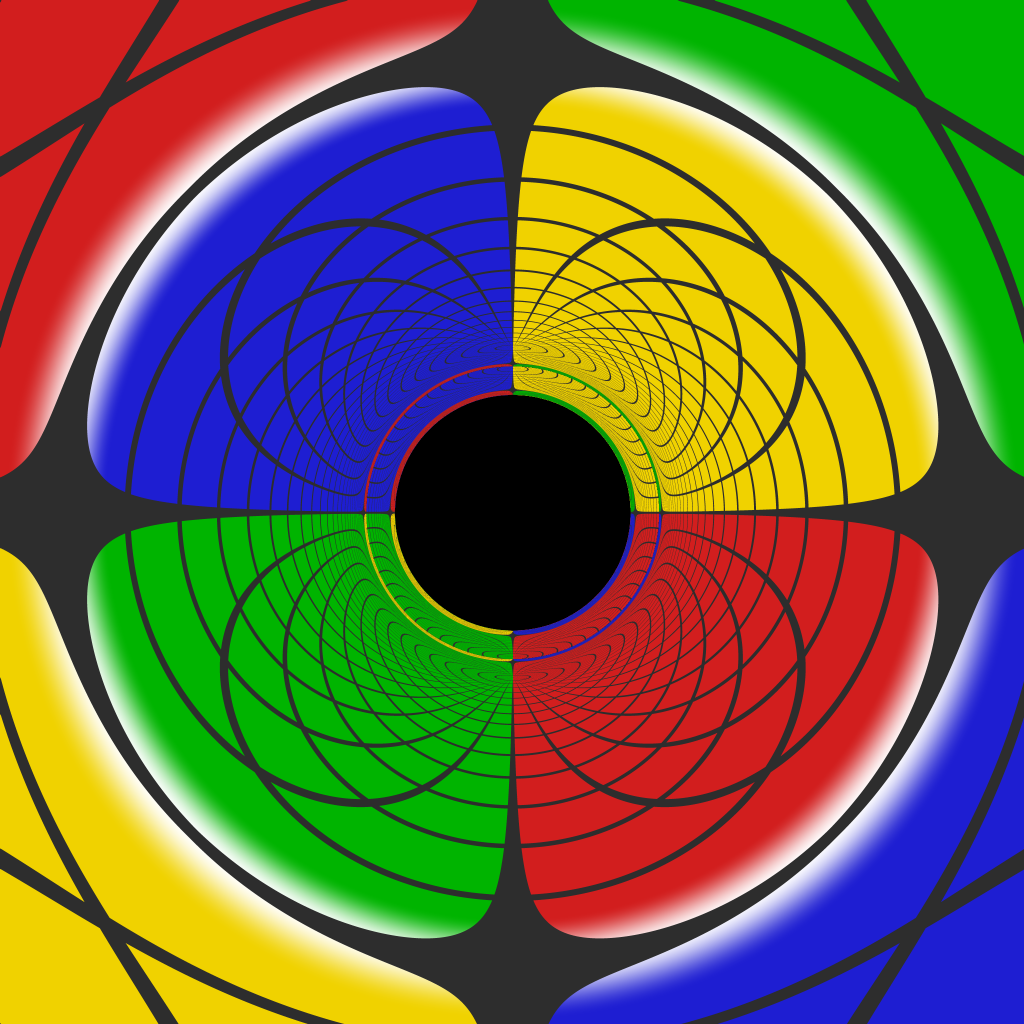} &
    \includegraphics[width=0.24\textwidth]{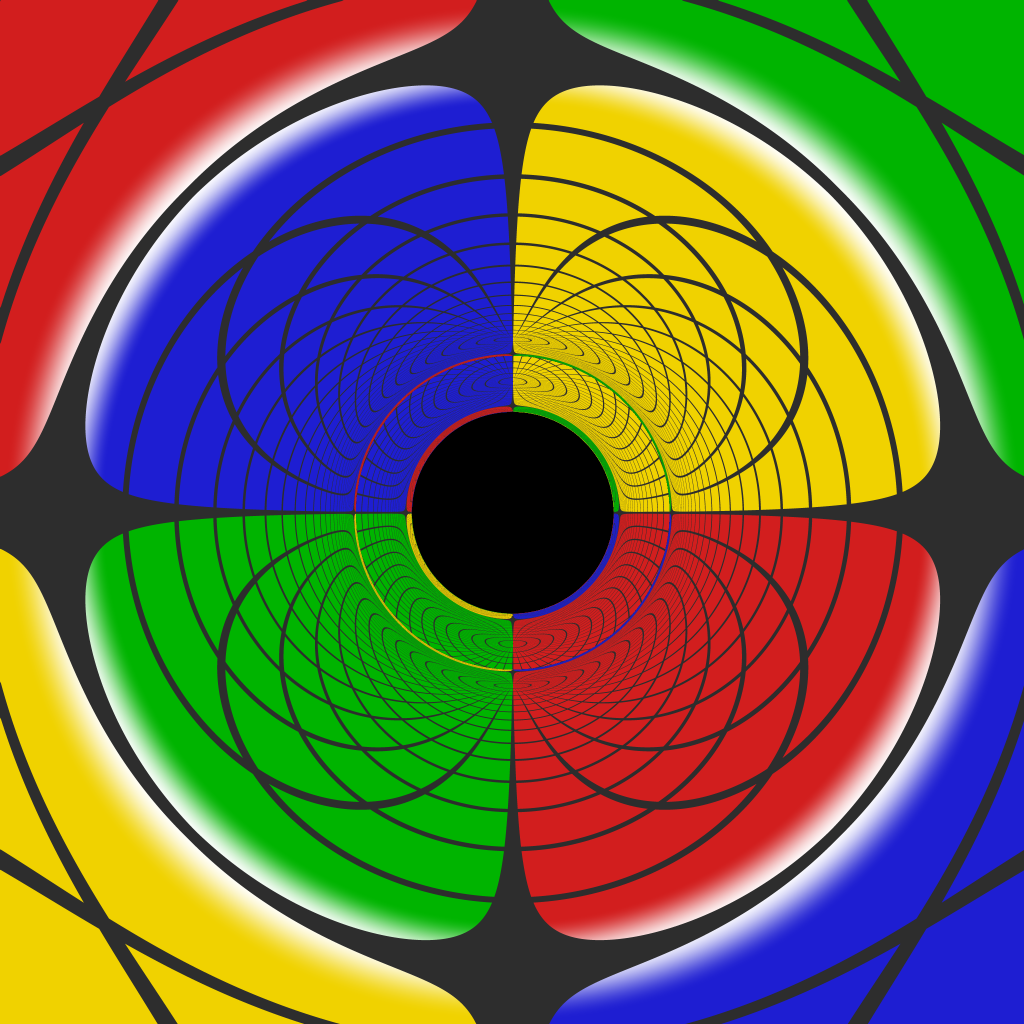} &
    \includegraphics[width=0.24\textwidth]{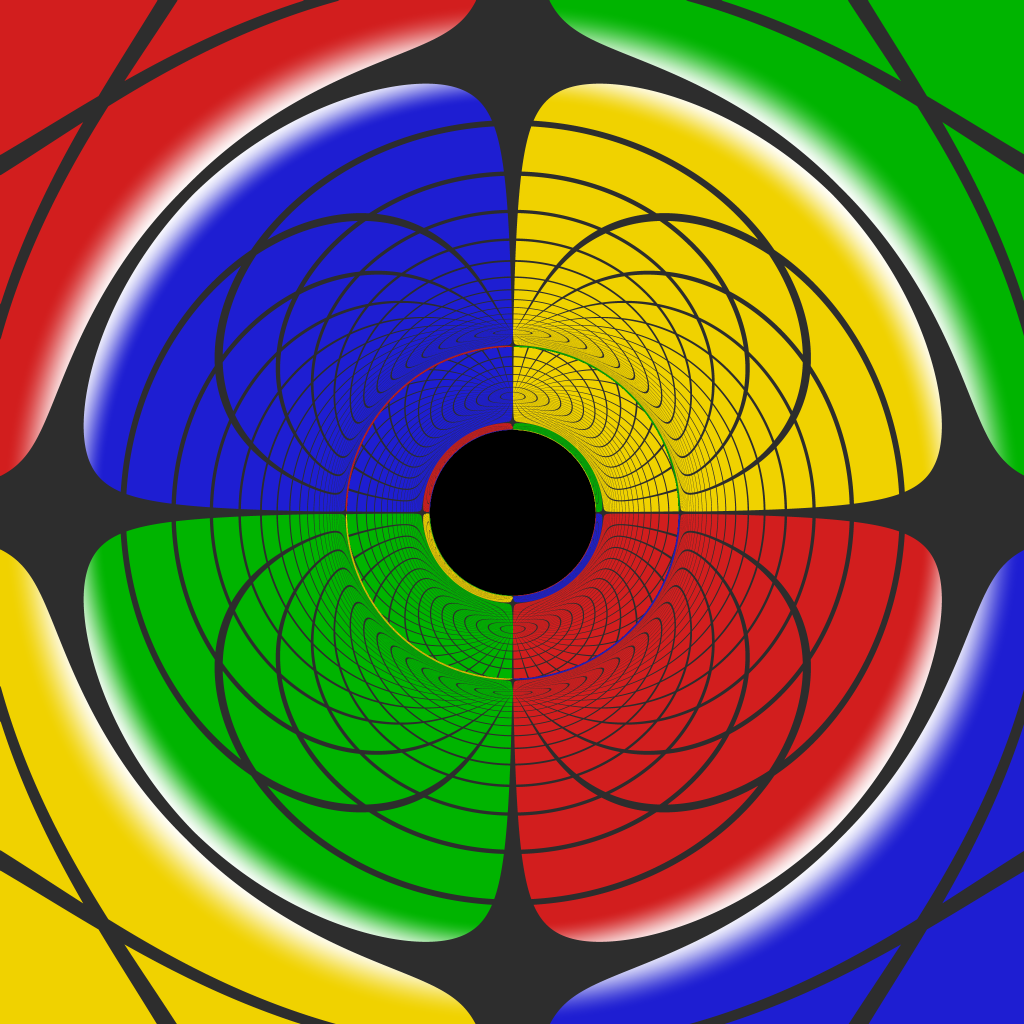}\\[2pt]
    {\small (a) $\alpha/M^2=-0.25$} & {\small (b) $\alpha/M^2=0.25$}
      & {\small (c) $\alpha/M^2=0.5$} & {\small (d) $\alpha/M^2=0.75$}\\
  \end{tabular}
  \caption{
  The two polarization shadows put together in one image, one panel per coupling:
  (a)~$\alpha/M^2=-0.25$, (b)~$0.25$, (c)~$0.5$, (d)~$0.75$.
  At each pixel we keep the ray that arrives later (the one with the larger travel time $t$),
  so the black disk marks where both polarizations fall into the hole,
  while the surrounding colored ring is reached by only one;
  the ring widens as $\alpha$ grows.}
  \label{fig06}
\end{figure}

All wave observables split in the same way, for one physical reason:
the two linear polarizations propagate in reciprocal optical metrics,
$\mathcal{W}^{\mathrm{PPM}}=1/\mathcal{W}^{\mathrm{PPL}}$ (Eq.~\eqref{eq05}).
Every quantity anchored to the photon sphere therefore follows a single function of the coupling, $\bc^{\pm}(\alpha)$:
degenerate at $\alpha=0$,
and splitting into a contracting PPM branch and an expanding PPL branch as $|\alpha|$ grows.
The same $\bc^{\pm}(\alpha)$ that sets the absorption threshold also sets the shadow edges, the glory spacing, and the photon-ring radius.
That coordination is itself the signature: a single achromatic split, oppositely signed for the two branches and shared by shadow, ring, glory, and absorption, is hard to reproduce with spin, inclination, or foreground.
Figures~\ref{fig02}--\ref{fig08} show, in order, the greybody factors, absorption spectra, the double shadow, the differential cross section, and the backward birefringence.

\paragraph{Greybody factors.} The modulus of the same matrix gives transmission for each mode,
the greybody factor $\Gamma^{\pm}_\ell=1-|S^{\pm}_\ell|^2$, the probability that part of the wave is absorbed by the barrier rather than reflected.
Figure~\ref{fig02} follows the three lowest partial waves, which dominate the low-frequency absorption.
As $\omega$ climbs past the top of the potential barrier,
each mode turns on in a smooth sigmoid step centered at $M\omega_{1/2}\simeq(\ell+\tfrac12)/(\bc/M)$ ($0.29$, $0.48$, $0.67$ for $\ell=1,2,3$ at $\alpha=0$), so higher $\ell$ turns on at higher frequency.
The coupling shifts each threshold the same way it shifts the phase:
PPM (the smaller $\bc$) turns on later, PPL (the larger $\bc$) earlier,
and the two thresholds separate as $|\alpha|$ grows until,
at $\alpha/M^2=0.75$,
their ratio reaches the capture-radius ratio $\bc^{\mathrm{PPL}}/\bc^{\mathrm{PPM}}\simeq2$.
The polarization information carried by the phase in $\arg S_\ell$ thus reappears in $|S_\ell|$ as a shift of the transmission edge.
Since $\Gamma_\ell$ uses only the magnitude,
it is independent of the matching basis and is the most robust of all wave observables.

\paragraph{Absorption.} The absorption cross sections $\sigma^{\pm}_{\mathrm{abs}}(\omega)$ oscillate around the geometric capture areas $\pi(\bc^{\pm})^2$ and approach this value at high frequency;
these limiting values are shown in Figure~\ref{fig03}(a,b) as dashed lines of the same color.
Reaching this limit (computed-to-geometric ratio $1.003$ at $M\omega=5$) both verifies the chain end to end and shows how the capture cross section varies with $\alpha$ via $\bc^{\pm}(\alpha)$:
at $\alpha/M^2=0.75$ the two high-frequency limits differ by a factor $(\bc^{\mathrm{PPL}}/\bc^{\mathrm{PPM}})^2=4.0$.
The regular ripple about the limit is the diffraction pattern set by the photon-sphere orbit~\cite{ref19},
whose period $\sim1/\bc^{\pm}$ again carries the coupling.
Since the absorption channel uses only the modulus, it is the most robust observable,
and we use it as a reference for separating matching errors;
its behavior is consistent with the partial-wave behavior previously observed in other contexts~\cite{ref22, ref23}.

\paragraph{Double shadow.}
The high-frequency limit of the absorption channel---the geometric capture area $\pi \bc^2$---is precisely the critical curve defining the shadow;
the polarization splitting of $\bc^{\pm}(\alpha)$ in absorption therefore reappears directly as a double shadow.
Each polarization has its own critical curve~\cite{ref30, ref31, ref32, ref33}:
as $\alpha$ grows, the PPM shadow contracts and the PPL shadow expands relative to the Schwarzschild value $\bc=3\sqrt{3}\,M$.
To visualize this we use a four-color celestial-sphere renderer---a standard technique for strong-lensing images~\cite{ref34, ref35}, following Bohn \emph{et al.}:
an observer at $r_{\mathrm{obs}}=100\,M$ views a background sphere at $r_{\mathrm{cel}}$ whose four quadrants carry different colors, an angular grid, and a faint white reference spot at the optical axis;
the shadow is the set of viewing directions whose backward-traced rays end on the horizon.
For quantitative work---the shadow's angular size and the gap between the two polarization edges---we use an orthographic impact-parameter camera
($b=\mathrm{pixel}\times M$, the correct far-observer limit) with half-width $20\,M$,
whose central pixel maps to the point on the celestial sphere directly behind the hole;
perspective (field-of-view) cameras are reserved for close-ups.
The reference spot falls on a continuous Einstein ring, which validates the lensing map.
Figure~\ref{fig04} shows the setup for a flat ($M=0$) background,
where the quadrants meet at the analytic crossing $\Phi_{\mathrm{flat}}(b)=\arccos(b/r_{\mathrm{obs}})+\arccos(b/r_{\mathrm{cel}})$ and there is no shadow,
and for Schwarzschild, where the central shadow and bright Einstein ring appear;
getting the flat map right is the first check of the renderer.
We cross-validated the rendering results using three independent geodesic calculation algorithms:
the reference images use an adaptive Runge--Kutta integrator,
and the same scene is reproduced by a GPU ray tracer in the style of \textsc{Mahakala}~\cite{ref36} and by the \textsc{gyoto} code~\cite{ref37},
both standard general-relativistic ray tracers~\cite{ref38};
the three tracers assign the same four-color class to each pixel,
with residual differences confined to the anti-aliased boundary of the critical curve.
Figure~\ref{fig05} shows the resulting images for two polarization scenarios and several values of $\alpha$.
As shown in Figure~\ref{fig03}(c), the two critical radii $\bc^{\pm}(\alpha)$ diverge smoothly from a common Schwarzschild value of $3\sqrt3\,M$, the PPM system contracts, while the PPL system expands.
Figure~\ref{fig06} puts both shadows into one image for each coupling.
At each pixel we keep the later-arriving ray,
so the black core is where both polarizations are captured and the colored ring around it is reached by only one;
the ring widens as $\alpha$ grows.

\section{Differential scattering and birefringence}\label{sec05}

\begin{figure}[t]
  \centering\includegraphics[width=\textwidth]{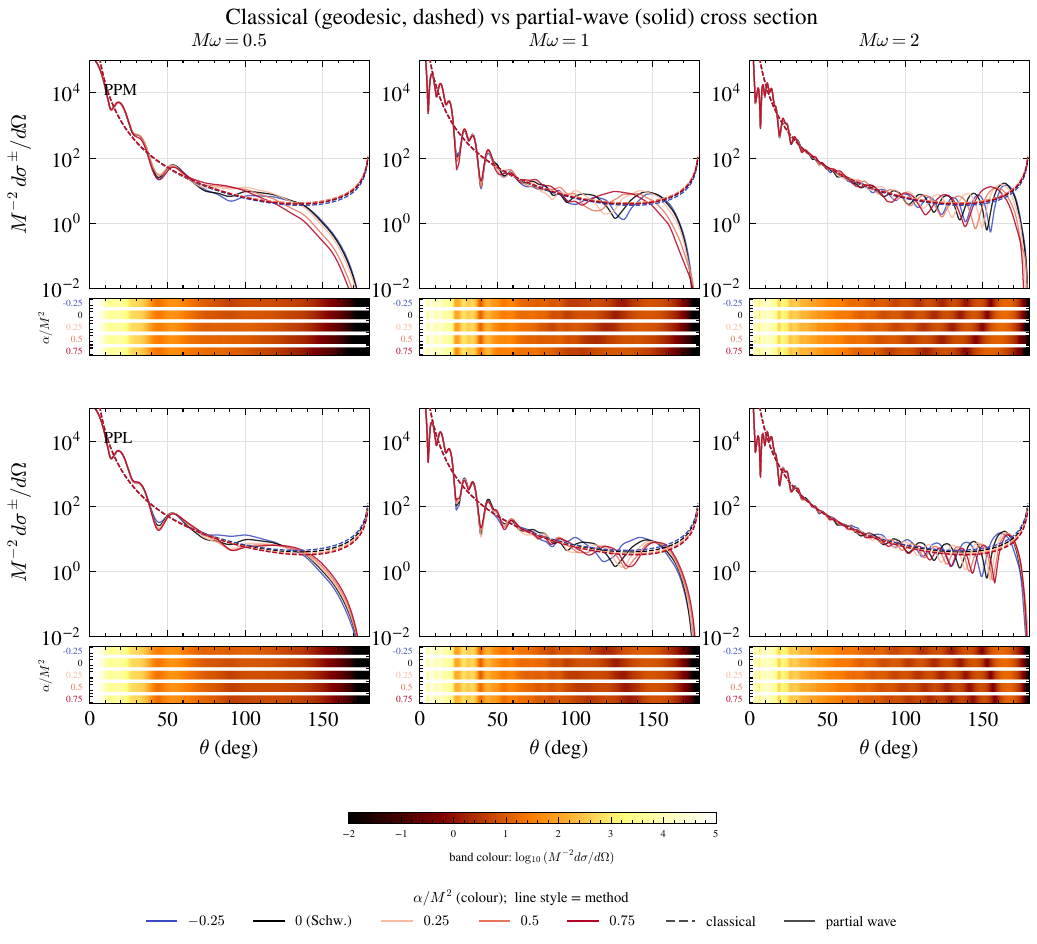}
  \caption{
  Classical (geodesic, dashed) versus partial-wave (solid) differential cross section for PPM (top) and PPL (bottom) at $M\omega=0.5,1,2$ (columns),
  for the five couplings $\alpha/M^2\in\{-0.25,0,0.25,0.5,0.75\}$ (color; $\alpha=0$ black).
  The wave result oscillates around the frequency-independent classical geodesic cross section of Eq.~\eqref{eq10} and converges toward it as $M\omega$ grows, away from the forward and backward limits;
  the backward rise is the orbiting/glory region, and the forward $\theta^{-4}$ tail is the Newtonian divergence.
  The coupling shifts the classical curve (through $\bc^{\pm}$) and the wave fringes together.
  The strips below each panel encode $M^{-2}\,\dd\sigma/\dd\Omega$ versus $\theta$ on the color scale shown, one row per $\alpha$ (row label colored to match its curve),
  making the forward peak and the fringe pattern immediately visible.}
  \label{fig07}
\end{figure}

\paragraph{Differential scattering and the glory.}
The differential cross section shows the forward rise and the backward glory oscillations.
At $\alpha=0$ and $M\omega=2$ the glory peak sits at $163.4^\circ$;
our computed peak agrees with Crispino \emph{et al.}~\cite{ref17} to within $0.7^\circ$.
For $\alpha\neq0$ the two polarizations show different glory patterns (Fig.~\ref{fig07}).
The glory arises from backward interference of rays that skirt the photon sphere~\cite{ref39, ref40, ref41},
so its fringe spacing is set by the critical impact parameter;
as $\bc^{\pm}(\alpha)$ splits, the PPM and PPL glories move in opposite directions, the wave-optical counterpart of the double shadow.
Each parity crosses its own barrier, so its fringe spacing follows $\bc^{\mathrm{PPM}}$ or $\bc^{\mathrm{PPL}}$.
At $\theta=\pi$ the $\alpha=0$ cross section drops to the electromagnetic backward zero;
once $\alpha\neq0$, the parity difference fills this zero, which is the basis of the birefringence observable below.
This backward behavior is characteristic of spin 1 and distinguishes photons from the scalar case~\cite{ref29}:
the glory amplitude takes the form $J_{2s}(\omega b_g\sin\theta)$, so for a scalar ($s=0$), $J_0$ reaches a peak at $\theta=\pi$ (a bright backward glory),
whereas for photons ($s=1$), $J_2$ is zero at that point, resulting in a zero rear component and shifting the halo maximum off-axis to $\simeq163^\circ$;
whereas the forward $16M^2/\theta^4$ tail is spin-independent and present in both cases.

\paragraph{Classical cross section.}
As a benchmark for geometric optics, we have followed the scalar scattering approach proposed by Lima-Junior \emph{et al.}~\cite{ref29} to calculate the classical sections corresponding to the zero-curvature geodesics of each optical metric.
The deflection angle for impact parameter $b$ is
\begin{equation}
  \chi^{\pm}(b)=2\!\int_{r_0}^{\infty}\!\frac{b\,\mathcal{W}^{\pm}(r)/r^2}
  {\sqrt{1-b^2 V^{\pm}_{\mathrm{geo}}(r)}}\,\dd r-\pi,
  \label{eq10}
\end{equation}
where $r_0$ is the outer root of $b^2 V^{\pm}_{\mathrm{geo}}=1$ and $V^{\pm}_{\mathrm{geo}}$ is from Eq.~\eqref{eq07};
the classical cross section is $\dd\sigma^{\pm}_{\mathrm{cl}}/\dd\Omega=\sum_b b\,/(\sin\theta\,|\dd\chi/\dd b|)$,
summed over the geodesic branches that scatter into $\theta$.
The partial wave result oscillates about this classical curve and follows it except near the forward and backward limits (Fig.~\ref{fig07}).
As $b\to\bc^{\pm}$ the deflection $\chi^{\pm}$ diverges (the ray orbits the photon sphere of its own polarization), so the classical curves for PPM and PPL blow up at their respective $\bc^{\pm}(\alpha)$;
this is the geometric-optics origin of the polarization-split glory.

\paragraph{Scattering cross section.}
The integrated scattering cross section $\sigma^{\pm}_{\mathrm{sc}}$ of Eq.~\eqref{eq09} is dominated by small-angle deflection and is the least sensitive to the coupling.
Both $\sigma_{\mathrm{sc}}$ and the angular integral of $\dd\sigma/\dd\Omega$ are rendered finite by the same $\ell$ truncation
(the untruncated sums diverge in the forward direction),
so with a common $\ell_{\max}$ they must agree identically up to the numerical treatment of the angular sum;
we use this as an internal consistency check on the amplitude construction of Eqs.~\eqref{eq10a}--\eqref{eq10c}
(agreement at the $10^{-3}$ level over the resolved angular range; Sec.~\ref{sec03num})
rather than as a probe of $\alpha$.

\begin{figure}[t]
  \centering
  \setlength{\tabcolsep}{4pt}%
  \begin{tabular}{@{}cc@{}}
    \includegraphics[width=0.49\textwidth]{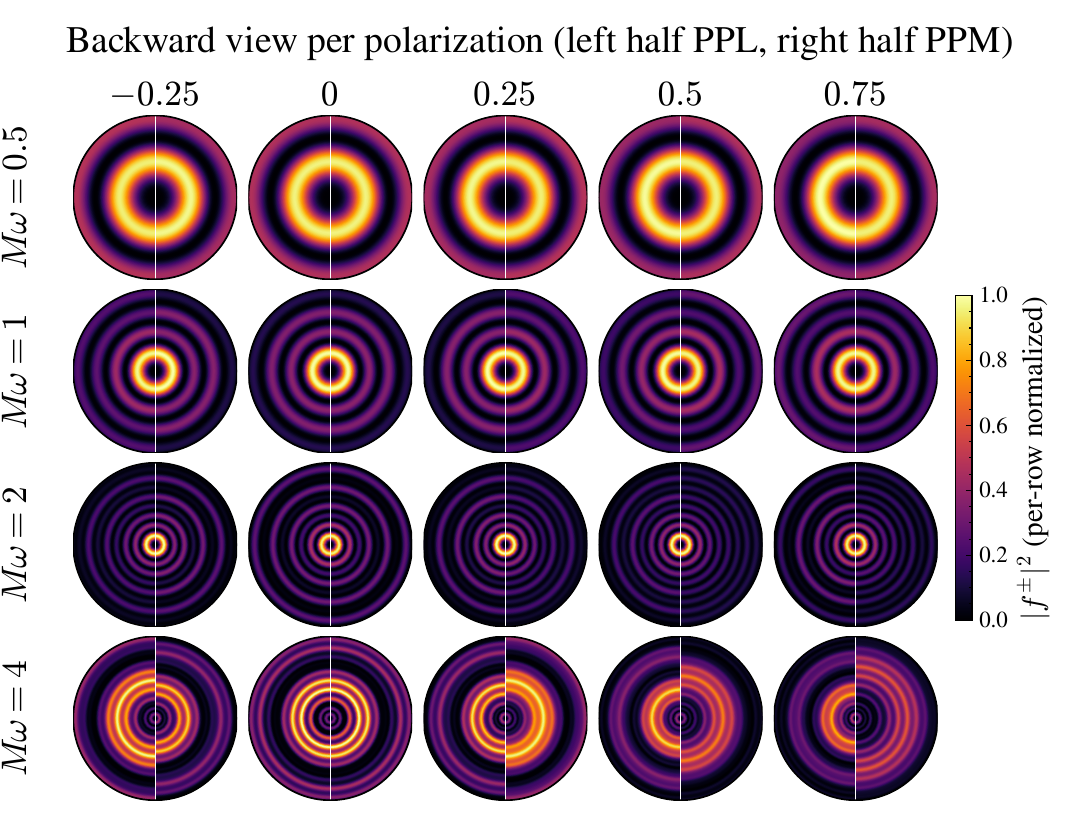} &
    \includegraphics[width=0.49\textwidth]{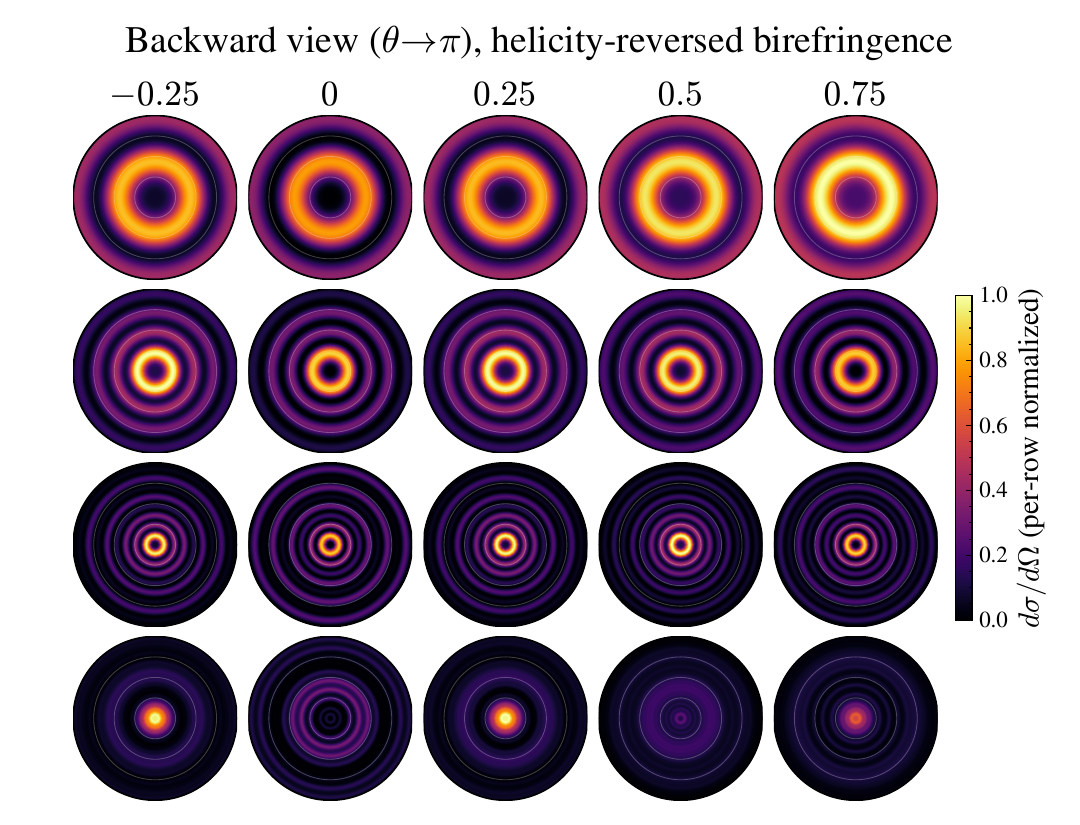}\\[1pt]
    {\small (a) per polarization} & {\small (b) polarization-summed}\\
  \end{tabular}
  \caption{
  Backward view ($\theta\to\pi$, radius $180^\circ-\theta$ up to $20^\circ$ from the antipode),
  rows $M\omega=0.5,1,2,4$ and columns $\alpha/M^2=-0.25,0,0.25,0.5,0.75$ (each row normalized to its own maximum).
  \textbf{(a)} Per-polarization: each disk's left half is PPL and right half PPM (the single-parity intensity $|f^{\pm}|^2$).
  For a single parity $g=0$, so each half is dark-centered (the helicity-preserving amplitude vanishes at $\theta=\pi$) with a glory ring;
  the PPL/PPM ring radii differ because $\bc^{\mathrm{PPL}}\neq\bc^{\mathrm{PPM}}$,
  and the left/right asymmetry flips between $\alpha$ and $-\alpha$.
  The backward spot is absent here; it is a parity-difference effect, seen only in (b).
  \textbf{(b)} Polarization-summed differential cross section, the analog of Fig.~2 of Crispino \emph{et al.}~\cite{ref18}.
  The helicity-preserving amplitude vanishes identically at $\theta=\pi$,
  so the exact backward flux is purely the helicity-reversed amplitude generated by the parity splitting:
  at $\alpha=0$ the center is dark (electromagnetic backward zero,
  surrounded by glory rings) at every frequency, while for $\alpha\neq0$ a bright spot lights up the antipode,
  faint at low frequency and growing with $M\omega$ --
  the geometric-optics image of the wave-optics birefringence.}
  \label{fig08}
\end{figure}

\begin{figure}[t]
  \centering\includegraphics[width=\linewidth]{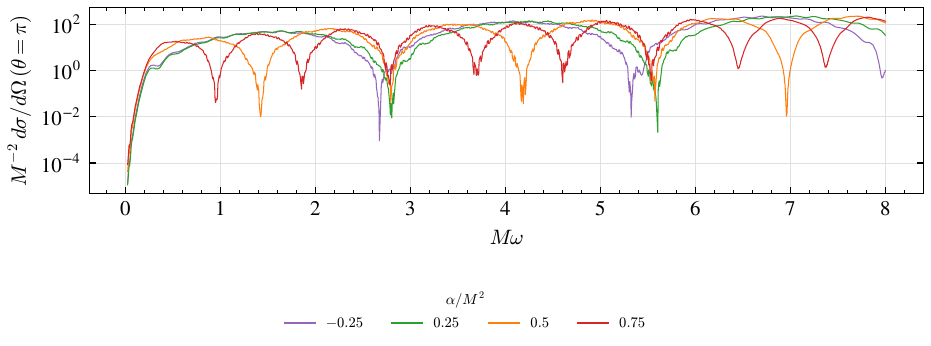}
  \caption{
  Backward birefringence, quantitatively:
  the exact-backscattering differential cross section $M^{-2}\,\dd\sigma/\dd\Omega|_{\theta=\pi}=M^{-2}|g(\pi)|^{2}$ versus $M\omega$
  for the five couplings (color; $\alpha=0$ in black, identically zero and shown only as the baseline).
  The signal oscillates under a slowly rising, glory-scale envelope,
  with deep nulls from the beating of the two parity glories:
  the null spacing $\Delta(M\omega)\simeq\pi M/(\bc^{\mathrm{PPL}}-\bc^{\mathrm{PPM}})$ shrinks as the coupling grows,
  making the backward spectrum a direct interferometric readout of the critical-curve splitting.}
  \label{fig08b}
\end{figure}

\paragraph{Birefringence.}
Figure~\ref{fig08b} quantifies the backward signal.
The exact-backscattering cross section $|g(\pi)|^2$ is identically zero at $\alpha=0$
(numerically $<10^{-12}$ over the whole frequency range);
for $\alpha\neq0$ it oscillates with frequency under a slowly rising, glory-scale envelope
(reaching $1.1\times10^{2}\,M^2$ at $M\omega=5$ and of order $2\times10^{2}\,M^2$ near $M\omega=8$ at $\alpha/M^2=0.75$),
punctuated by deep interference nulls.
The envelope carries little information about $\alpha$,
but the nulls do:
they arise from the beating of the two parity glories,
so their spacing is set by the splitting of the critical impact parameters,
$\Delta(M\omega)_{\mathrm{null}}\simeq\pi M/(\bc^{\mathrm{PPL}}-\bc^{\mathrm{PPM}})$.
The measured spacings, $0.92$, $1.40$, and $2.77$ at $\alpha/M^2=0.75$, $0.5$, and $0.25$,
match this estimate ($0.97$, $1.43$, $2.80$) to a few percent:
the backward spectrum is a direct interferometric readout of $\bc^{\mathrm{PPL}}-\bc^{\mathrm{PPM}}$, hence of $\alpha$.
Of all wave-optical observables this is the purest probe of the coupling, since it extracts only the parity difference.
The backward zero has a clear physical origin: backward scattering reverses the helicity of photons;
for a field with spin 1, the amplitude for which the helicity \emph{remains unchanged} at $\theta=\pi$ is always zero,
so only the amplitude for which the helicity \emph{is reversed} survives at this point, and this amplitude is given by the parity difference $S^{M}-S^{E}$.
In the minimal theory, these two parity types are degenerate ($S^{E}=S^{M}$), so the backward flux is exactly zero; this is a symmetry of the Schwarzschild electromagnetic problem,
and thus any backward signal can directly measure the extent to which the coupling breaks this symmetry.
This is identical to the mechanism by which a black hole's charge $Q$ causes emission in the backward direction~\cite{ref18}, where $\alpha$ here plays the role of $Q$.
Following polarization separation (Fig.~\ref{fig08}(a)), the center of each inverted image, whether PPM or PPL, is a dark region;
when viewed in isolation, neither image reveals this spot, as each individual parity satisfies $g=0$, while the glories of PPM and PPL are located at the separation radii $\bc^{\pm}$ respectively.
This bright spot appears only in the composite image (Fig.~\ref{fig08}(b)); when coupling causes the two polarization states to separate, the spot grows larger as the parity difference increases.

\section{Strong-field optics: photon ring and trajectories}\label{sec06}

\begin{figure}[t]
  \centering\includegraphics[width=\textwidth]{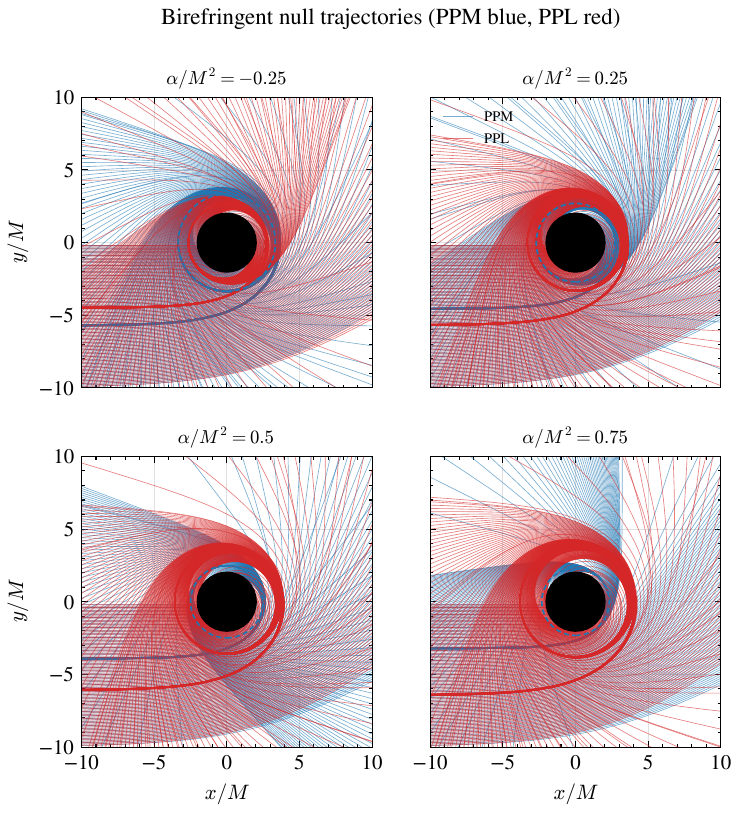}
  \caption{
  Birefringent null trajectories:
  a parallel beam incident horizontally from the far lower left (entry heights $y/M\in[-10,0]$),
  sampled in three nested sets of increasing density --
  a wide uniform beam plus two finer bands of width $0.1\,M$ and $10^{-3}\,M$ centered on each polarization's critical launch value --
  so the deflection is resolved both far out and right at the photon sphere,
  for $\alpha/M^2=-0.25,0.25,0.5,0.75$ (panels).
  The two polarizations (PPM blue, PPL red) follow distinct optical metrics and split,
  the sign of the splitting reversing between $\alpha<0$ and $\alpha>0$;
  rays with $b<\bc^{\pm}$ are captured by the hole (black disk,
  the horizon) and the rest are deflected by different amounts --
  the geometric-optics birefringence.
  Dashed circles mark the unstable photon spheres $r_{\mathrm{ph}}^{\pm}$,
  onto which the near-critical rays pile up.}
  \label{fig09}
\end{figure}

The double shadow of Sec.~\ref{sec04} is the static face of the photon-sphere splitting.
The photon ring and the null geodesic carry the same splitting into the sharp,
finely imaged substructure of the strong-field region.

The photon ring exhibits the same splitting~\cite{ref42, ref43}.
Successive subrings are images of light that completed one extra half-orbit around the photon sphere;
each half-orbit demagnifies them by $e^{-\gamma^{\pm}}$,
with $\gamma^{\pm}$ the Lyapunov exponent of the unstable orbit~\cite{ref44},
and they accumulate onto the critical curves $\bc^{\pm}$.
Table~\ref{tab01} lists $\gamma^{\pm}(\alpha)$ computed from the geometric-optics potential of Eq.~\eqref{eq07}
($\gamma=\pi\sqrt{-\tfrac12\,\partial_u^2[V^{\pm}_{\mathrm{geo}}]}$ at $u=1/r_{\mathrm{ph}}$, with $u\equiv1/r$; the Schwarzschild value is $\pi$).
The splitting is strongly asymmetric between the branches:
the contracting PPM ring stack fades quickly
($\gamma^{\mathrm{PPM}}=1.70\pi$ at $\alpha/M^2=0.5$ and $2.65\pi$ at $0.75$,
so each PPM subring is $\sim200$--$4000\times$ fainter than the last),
while the expanding PPL stack stays close to the Schwarzschild decay ($\gamma^{\mathrm{PPL}}\simeq0.92\pi$).
Because $\bc^{\mathrm{PPM}}$ and $\bc^{\mathrm{PPL}}$ move in opposite directions,
the two polarizations form an interlaced ring system whose spacing is set by $\alpha$.
This polarized substructure is the geometric-optics counterpart of the wave-optics birefringence,
and unlike the fuzzy shadow edge it is sharp and largely free of astrophysical foregrounds.
For a hypothetically resolvable ring,
the interferometric visibility would oscillate with a radial period set by the ring diameter $\sim2\bc$~\cite{ref45};
polarization-resolved visibilities would then show two slightly different periods,
whose fractional offset equals $(\bc^{\mathrm{PPL}}-\bc^{\mathrm{PPM}})/\bc\approx0.83\,\alpha/M^2$ (Sec.~\ref{sec07}),
so the period shift reads off $\bc^{\mathrm{PPL}}-\bc^{\mathrm{PPM}}$ directly;
we recall from Sec.~\ref{sec02} that for the black holes whose rings could actually be resolved this splitting is unobservably small.
The null geodesic lines plotted using the two optical metrics directly reveal the phenomenon of birefringent bending (Fig.~\ref{fig09}): two polarized light rays incident in the same manner are either captured or escape at different critical angles.

In the geometric-optics reduction used here, the light-cone factor $\mathcal{W}^{\pm}$ rescales only the angular part of the optical metric and leaves the radial cone $\dd r/\dd t=\pm f(r)$ unchanged.
Consequently, for both polarization states, the effective photon horizon lies at the Schwarzschild horizon $r=2M$;
captured light rays are absorbed at this point rather than being re-emitted, which is consistent with the pure incidence condition we imposed at $r=2M$ in the wave problem (see Sec.~\ref{sec03}).
The full non-minimal coupling could in principle tilt the radial cone too,
making one polarization superluminal relative to the background metric and moving its effective horizon off $r=2M$,
so that photons might escape from inside the geometric horizon~\cite{ref01, ref02};
capturing that would need the full polarization-dependent dispersion relation rather than the angular-only reduction we use here,
so we leave it aside.

\section{Observational implications}\label{sec07}

The scale argument of Sec.~\ref{sec02} settles which black holes these observables could apply to.
For supermassive ones it is decisive:
with $\alpha$ bounded at the few-$\mathrm{m}^2$ level by laboratory birefringence searches~\cite{ref49},
the ratio $\alpha/M^2$ for M87$^*$ cannot exceed $\sim10^{-26}$,
so the Weyl coupling leaves no observable imprint on the shadows, photon rings, or polarized images accessible to horizon-scale interferometry~\cite{ref08, ref09, ref10}, now or in the foreseeable future.
Shadow-size measurements of M87$^*$ and Sgr~A$^*$ have been translated into constraints on other non-Kerr and non-minimal extensions~\cite{ref46, ref47, ref48},
and formally our $\bc^{\pm}(\alpha)$ is the quantity such analyses bound;
but for this theory the accessible precision falls short of the required one by more than twenty orders of magnitude.

Order-unity $\alpha/M^2$ survives only at the very bottom of the black-hole mass range:
for $M\lesssim10^{15}\,\mathrm{g}$ the Standard-Model (Drummond--Hathrell) coefficient alone gives $\alpha/M^2\gtrsim1$ (Sec.~\ref{sec02}),
so for a primordial black hole this light---should such objects exist and be detected~\cite{ref50}---the birefringence computed here would be an order-one property of its electromagnetic environment.
We put this forward as the one regime the laboratory bounds leave open, not as an expectation:
no primordial black hole has been observed, the lightest ones have already evaporated,
and the abundance of those just above the evaporation threshold is itself constrained~\cite{ref50}.
In that regime our observables split into two groups.
The shadow of such an object is far too small to be resolved by any imaging,
so the double shadow and the polarization-split photon ring of Secs.~\ref{sec04} and~\ref{sec06} serve as theory diagnostics and as the geometric-optics anchors of the wave calculation, not as imaging targets.
Any actual signal would come from the wave-optics quantities:
the polarization-resolved greybody factors and absorption cross sections of Sec.~\ref{sec04}---which,
for black holes near the end of their Hawking evaporation, directly shape the polarization content of the emitted photon spectrum---the
differential cross sections and glory structure of Sec.~\ref{sec05},
and the backward birefringence,
which is protected by parity, vanishes identically at $\alpha=0$,
and whose interference-null spacing reads off the splitting $\bc^{\mathrm{PPL}}-\bc^{\mathrm{PPM}}$ directly.
The near-linear relation between splitting and coupling,
$(\bc^{\mathrm{PPL}}-\bc^{\mathrm{PPM}})/\bc\approx0.83\,\alpha/M^2$ (reaching $0.62$ at $\alpha/M^2=0.75$),
converts any measured polarization asymmetry into a value of $\alpha/M^2$ and,
with an independent mass estimate, into $\alpha$ itself.
We deliberately stop short of a detection forecast:
the relevant fluxes depend on the unknown abundance and environment of such objects.
What this paper supplies is the theory-side input such a search would need,
computed for both polarizations from the wave equation through to the geometric-optics limit.

\section{Conclusions}\label{sec08}

For photons coupled to the Weyl tensor, and for two polarization states,
we have calculated the wave-optical and geometric-optical observables of a Schwarzschild black hole.
This coupling causes the black hole to exhibit birefringence:
it separates the phase shift, absorption and differential cross-section, producing a backscattering signal that is exactly zero in the absence of this coupling,
and casting a double shadow featuring a polarization-dependent photon ring.
The wave-optical splitting acquires quantitative meaning through the Riccati--Hankel extraction of the scattering matrix;
we have identified which observables that accuracy governs (the phase-sensitive cross sections, rather than the modulus-only absorption),
and verified the geometric-optics chain with three independent ray tracers,
whose four-color images agree away from the anti-aliased boundary of the critical curve.
The cleanest signature is the backward birefringence, which is protected by parity, is exactly zero in the minimal theory, and whose interference-null spacing reads off the splitting $\bc^{\mathrm{PPL}}-\bc^{\mathrm{PPM}}$ directly.
Natural next steps are a rotating (Kerr) background, a frequency-dependent coupling,
and the application of the polarization-resolved greybody factors to the Hawking emission of light primordial black holes,
which would be the only objects light enough for the couplings displayed here to escape the laboratory bounds (Sec.~\ref{sec02}).

\acknowledgments
This work was supported by the Zhejiang normal university Doctorial research fund Contract No. ZC302924005.

\bibliographystyle{JHEP}
\bibliography{refs}

\end{document}